\documentclass[prb,aps,amssymb,showpacs,twocolumn]{revtex4-1}
\usepackage{amsmath}
\usepackage{amssymb}
\usepackage{amsthm}
\usepackage{amsfonts}
\usepackage{listings}
\usepackage{enumerate}
\usepackage{latexsym}

\usepackage{psfrag}

\usepackage{bm}
\usepackage{graphicx}
\usepackage{subfigure}

\newcommand{\beq}{\begin{equation}}
\newcommand{\eneq}{\end{equation}}
\def\ie{{\it i.e.},\ }

\input{epsf}

\begin{document}

\tolerance 10000

\newcommand{\vk}{{\bf k}}

%\draft

\title{Series of Abelian and Non-Abelian States in $C>1$ Fractional Chern Insulators}

\author{A. Sterdyniak$^1$}
\author{C. Repellin$^1$}
\author{B. Andrei Bernevig$^2$}
\author{N. Regnault$^{2,1}$}
\affiliation{$^1$ Laboratoire Pierre Aigrain, ENS and CNRS, 24 rue Lhomond, 75005 Paris, France\\
$^2$ Department of Physics, Princeton University, Princeton, NJ 08544}

\begin{abstract}
We report the observation of a new series of Abelian and non-Abelian topological states in fractional Chern insulators (FCI). The states appear at bosonic filling $\nu= k/(C+1)$ ($k, C$ integers) in several lattice models, in fractionally filled bands of Chern numbers $C\ge 1$ subject to on-site Hubbard interactions. We show strong evidence that the $k=1$ series is Abelian while the $k >1$ series is non-Abelian. The energy spectrum at both groundstate filling and upon the addition of quasiholes shows a low-lying manifold of states whose total degeneracy and counting matches, at the appropriate size, that of the Fractional Quantum Hall (FQH) $SU(C)$ (color) singlet  $k$-clustered states (including Halperin, non-Abelian spin singlet(NASS) states and their generalizations). The groundstate momenta are correctly predicted by the FQH to FCI lattice folding. However, the counting of FCI states also matches that of a spinless FQH series, preventing a clear identification just from the energy spectrum. The 
entanglement spectrum lends support to the identification of our states as $SU(C)$ color-singlets but offers new anomalies in the counting for $C>1$, possibly related to dislocations that call for the development of new counting rules  of these topological states. 
\end{abstract}

\date{\today}


\maketitle

\section{Introduction}

In his seminal paper\cite{haldane-1988PhRvL..61.2015H}, Haldane introduced the concept of a Chern insulator (CI). Classified by a non-zero Chern number of the occupied bands, this band insulator exhibits an integer Hall conductance akin to the integer quantum Hall effect but in the absence of an overall magnetic field. Recently, the role of strong interactions in CI has received strong  attention \cite{neupert-PhysRevLett.106.236804,sheng-Natcommun.2.389,regnault-PhysRevX.1.021014,Bernevig-2012PhysRevB.85.075128,Wu-2012PhysRevB.85.075116,Venderbos-PhysRevLett.108.126405,wang-PhysRevLett.107.146803,parameswaran-PhysRevB.85.241308,goerbig-2012epjb,Neupert-2011PhRvB..84p5107N,Neupert-2011PhRvB..84p5107N,Murthy-2011arXiv1108.5501M,Murthy-PhysRevB.86.195146} focused on realizing the FQH in flat bands akin to Landau levels but in zero field.  Most of these studies have focused on CI with fractionally filled bands of Chern number $C$ unity. These systems have been dubbed Fractional Chern Insulators (FCI). Strong numerical \cite{neupert-PhysRevLett.106.236804,sheng-Natcommun.2.389,Bernevig-2012PhysRevB.85.075128,Wu-2012PhysRevB.85.075116,Venderbos-PhysRevLett.108.126405} and analytical \cite{Wu-2012arXiv1206.5773W} evidences of Laughlin-like~\cite{laughlin-PhysRevLett.50.1395} states or more generally composite fermion~\cite{jain89prl199} states have been obtained in CIs in the presence of 2-body interaction \cite{Liu-2012arXiv1206.2626L,Lauchli-2012arXiv1207.6094L}. With $(k+1)$-body interactions, phases similar to the Moore-Read\cite{Moore1991362} (MR) or Read-Rezayi~\cite{read-PhysRevB.59.8084} (RR) states emerge~\cite{Bernevig-2012PhysRevB.85.075128,wang-PhysRevLett.108.126805,Wu-2012PhysRevB.85.075116}. 

A nice feature of CIs absent in the QH effect is the possibility to obtain a band with a higher Chern number, without introducing an additional degree of freedom (like the spin in the FQH) \ie $C>1$ models do not have any $SU(C)$ symmetry. Several simple one-body models of $C>1$ CIs with an almost flat band now exist \cite{wang-PhysRevB.84.241103,Barkeshli-2011arXiv1112.3311B,Wang-PhysRevB.86.201101,Trescher-PhysRevB.86.241111,Yang-PhysRevB.86.241112}. Ref.~\onlinecite{Barkeshli-2011arXiv1112.3311B} proposed, based on the Wannier approach \cite{qi-PhysRevLett.107.126803, Wu-2012arXiv1206.5773W, Yangle-WuFuturePaper}, that non-interacting CIs with $C>1$ are identical to the the quantum Hall effect \ie $C=1$ insulators, with particles carrying an extra $SU(C)$ internal degree of freedom\cite{yanglenewpapercomment}. It has then been assumed that this property should also hold true in the strong interaction regime. Moreover, several recent numerical works~\cite{Wang-PhysRevB.86.201101,Liu-PhysRevLett.109.186805, Grushin-PhysRevB.86.205125} have reported the observation of FQH-like phases at filling factor $\nu=\frac{1}{C+1}$ for bosonic systems and $\nu=\frac{1}{2C+1}$ for fermionic systems, and presented evidence (based only on the energy spectrum) that they are Abelian. While the possibility that these states have an internal hidden structure (such as  some $SU(C)$ or $Z_C$ degree of freedom) was raised in \cite{Wang-PhysRevB.86.201101,Liu-PhysRevLett.109.186805,Grushin-PhysRevB.86.205125}, no evidence for this scenario was provided, and the observed energy spectra could equally be well-explained by spinless states.

In this paper, we  pose the natural question of whether generalizations of the MR or RR states can emerge in FCI when $C>1$. In this article, we present strong indications that such phases are realized in FCI at filling factor $\nu=\frac{k}{C+1}$ when a fractionally filled CI band is subject to a $(k+1)$-body interaction. If the analogy between the FCI and the $SU(C) FQH$ is valid, we expect that this series to be related to the Halperin state~\cite{Halperin83} (for $k=1$), the non-Abelian spin singlet~\cite{Ardonne-PhysRevLett.82.5096} (NASS) (for $C=2$) states and their generalizations~\cite{Ardonne2001549} (for higher $k$ values). 

The signatures of the FQH-like phases in $C = 1$ FCIs appear in different manners~\cite{Bernevig-2012PhysRevB.85.075128},  the most simple of which are the groundstate (quasi-)degeneracy, the counting of the number of quasihole states and their momentum quantum numbers whose values can be obtained from the FQH-FCI mapping described in Ref.~\onlinecite{Bernevig-2012PhysRevB.85.075128}. Unfortunately, a Charge Density Wave (CDW) would exhibit similar counting~\cite{Bernevig-2012arXiv1204.5682B} in its energy spectra. The particle entanglement spectrum~\cite{li-08prl010504,sterdyniak-PhysRevLett.106.100405} is a better way to identify signatures of FQH phases\cite{regnault-PhysRevX.1.021014}, and also to distinguish between CDW and FCI. The previous studies~\cite{Wang-PhysRevB.86.201101,Liu-PhysRevLett.109.186805,Grushin-PhysRevB.86.205125} of $C>1$ FCIs have mostly focused on the matching between counting of groundstate or quasihole manifold degeneracy of the FCI and FQH. Even excluding the possibility of CDW 
phases, as we clearly show this is {\emph{not enough to prove}} that the physics of  $C > 1$ FCIs is related to the spinful FQH effect: we show that the counting of any of the previously observed states~\cite{Wang-PhysRevB.86.201101,Liu-PhysRevLett.109.186805,Grushin-PhysRevB.86.205125} as well as our new non-Abelian series can be deduced from a {\emph{spinless}} generalized Pauli exclusion principle~\cite{Haldane-PhysRevLett.67.937,bernevig-PhysRevLett.100.246802}. We show that the entanglement spectrum is able to rule out the option of spinless FQH states while providing indication (some yet non-understood properties disallow a clear proof) of colorful FQH states.  More recently, approaches~\cite{Wu-2012arXiv1206.5773W,Scaffidi-PhysRevLett.109.246805,Wu-PhysRevB.86.165129}, some based on the gauge-fixed unitary~\cite{Wu-2012arXiv1206.5773W} Wannier basis~\cite{qi-PhysRevLett.107.126803} were used to compute overlaps and to adiabatically continue between FCI and FQH states in $C=1$ FCIs, but these 
approaches are not yet applicable in the current $C>1$ case~\cite{yanglenewpapercomment}.

This article is organized as follows. In Sec.~\ref{sec:SUCFQH}, we give an overview of several model wave functions for the FQH effect with a $SU(C)$ internal degree of freedom. We explain how the degeneracy of these model states can be derived from a generalized Pauli principle. In Sec.~\ref{sec:numerical}, we present the numerical simulations for several FCI models with $C>1$, both for two-body and three-body interactions. We discuss how the energy spectra cannot discriminate between a spinless and a  $SU(C)$ spinful physics. In  Sec.~\ref{sec:pes}, we discuss which fingerprints should appear through the entanglement spectroscopy for a spinless and a $SU(C)$ spinful phase. Analyzing the numerical results using this technique, we provide multiple evidence of a $SU(C)$ spin structure within the $C>1$ FCIs.

\section{$SU(C)$ FQH model wave functions}\label{sec:SUCFQH}

It was proposed in Ref.~\onlinecite{Barkeshli-2011arXiv1112.3311B} that $C>1$ Chern insulators are analogous to $C = 1$ systems of particles with a $SU(C)$ internal degree of freedom. Thus, we first review candidate FQH systems with $SU(C)$ internal symmetry. There, the generalized $[m;n]$-Halperin wavefunctions are the main Abelian canditates. They are given by:    
\begin{equation}
 \Psi_{[m;n]}^{SU(C)} = \Phi_{\{m\}}^{\mathrm{intra}} \Phi_{\{n\}}^{\mathrm{inter}}\exp\left(-\frac{1}{4}\sum_{i=1}^C\sum_{k_i = 0}^{N_i} |z_{k_i}^{(i)}|^2 \right)
\label{halperin}
\end{equation}
where 
\begin{equation}
\Phi_{\{m\}}^{\mathrm{intra}} = \prod_{i=1}^C\prod_{k_i<l_i} (z_{k_i}^{(i)}-z_{k_i}^{(i)})^{m}
\label{halperin_intra}
\end{equation}
is the product of a Laughlin state for each component and 
\begin{equation}
\Phi_{\{n\}}^{\mathrm{inter}} = \prod_{i<j}^C\prod_{k_i = 1}^{N_i}\prod_{k_j = 1}^{N_j}(z_{k_i}^{(i)}-z_{k_j}^{(j)})^n
\label{halperin_inter}
\end{equation}
accounts for correlations between components. Here, $z_{k}^{(i)}$ is the complex position of the $k$-th particle of component $i$. The exponents $m$ and $n$ characterize the strength of the intra and inter component correlations respectively. The total filling factor is $\nu_{\mathrm{FQH}} = \frac{C}{m+(C-1)n}$ and the groundstate degeneracy on the torus is $d=(m-n)^{C-1}(m+(C-1)n)$. These states are $SU(C)$-singlets when $n = m - 1$. In our case, we consider $m=2, n=1$.

From the states in Eq.~\ref{halperin}, one can build a series of non-Abelian spin singlet states~\cite{Ardonne-PhysRevLett.82.5096} at filling $\nu_{\mathrm{FQH}}=\frac{Ck}{C(m-1)+1}$. This is done by dividing the particles into $k$ groups, writing a Halperin $[m;m-1]$ state for each group and then symmetrizing over the different groups. This procedure leads to:
\begin{eqnarray}
 \Psi_{m}^{SU(C),k} = \mathcal{S}\left[\prod_{i = 0}^{k-1}\Psi^{SU(C)}_{[m,m-1]}(z_{i\frac{N}{k}+1},,..,z_{(i+1)\frac{N}{k}})\right]
\end{eqnarray}
where $\mathcal{S}$ is the symmetrization operator. For $m=2,n=1$, these are the NASS states introduced by Ref.~\onlinecite{Ardonne-PhysRevLett.82.5096} for $C=2$ and generalized for $C=4$ in Refs.~\onlinecite{PhysRevLett.89.120401,PhysRevA.69.023612}.

The counting of excitations of the series above has been determined through generalized Pauli principles \cite{Estienne2012185,Ardonne-PhysRevLett.82.5096}. All spinless fermionic (bosonic) many-body wave functions of $N_e$ particles can be expressed as linear combinations of Fock states in the occupancy basis of the single-particle orbitals $m_{\lambda} = \mathcal{S}\left[\prod_i z_i^{\lambda_i}\right]$. Each Fock state can be labeled either by $\lambda$, a partition, or by the occupation number configuration $n(\lambda)=\{n_l(\lambda), l = N_{\Phi},...,0\}$, where $n_l(\lambda)$ is the number of times $l$ appears in $\lambda$ and $N_{\Phi}$ is the number of flux quanta. In the spinful case, the partition is replaced by a dressed partition that mixes momentum and spin. A dressed partition $(\lambda,\sigma)$ is given by $N_e$ entries $\lambda_i$ and a spin dressing $\sigma \in [1,2,\ldots C] $ which obey $\lambda_i>\lambda_{i+1}$ or $\lambda_i=\lambda_{i+1}$ and $\sigma_i \geq \sigma_{i+1}$. The number of 
groundstates and quasiholes states for a given value of $N_e$ and $N_\Phi$ is given by the number of dressed partitions $(k,r)_C$ admissible. Such a partition obeys the following conditions:
\begin{equation}
 \lambda_i - \lambda_{i+k} \geq r \mathrm{~or~} \lambda_i - \lambda_{i+k} = r - 1 \mathrm{~and~} \sigma_i < \sigma_{i+k}\label{pauli}
\end{equation}

\section{Numerical results}\label{sec:numerical}

In this article, we mostly focus on the slightly simplified pyrochlore lattice model introduced in ref~\onlinecite{Trescher-PhysRevB.86.241111}, consisting of $N$ Kagome layers coupled through intermediate triangular layers. For completeness, in the notation of Ref.~\onlinecite{Trescher-PhysRevB.86.241111}, we use parameters $t_1 = 1$, $t_{\bot}=-1.03$, $\lambda_1=0.83$ and $t_2=\lambda_2=0$ (i.e. we have discarded the second nearest neighbor hopping). This model gives a flat band with Chern number $C=N$. We also investigate other CI models: the triangular lattice model\cite{Wang-PhysRevB.86.201101} with $C=2$, the two orbitals on a  triangular lattice\cite{Yang-PhysRevB.86.241112} with $C=3$ and the $C$-Orbital on a square lattice model for $C\geq3$ \cite{Yang-PhysRevB.86.241112}.

We consider $N_e$ interacting bosons without any internal degree of freedom on a lattice with $N_x\times N_y$ unit cells and periodic boundary conditions. For the pyrochlore and the triangular lattice models, we use a $(k+1)$-body Hubbard interaction 
\begin{equation}
H_{{\rm int}, k}^{\rm pyro}=U\sum_{i} :\rho_i^{k+1} :\label{HintPyro}
\end{equation}
where $::$ denotes the normal ordering and the sum runs over all the sites. For the $C$-Orbital on square lattice and two orbitals on triangular lattice models, we use an isotropic on site interaction 
\begin{equation}
H_{{\rm int}, k}^{\rm C-orb}=U\sum_{i} :(\sum_j \rho_{i,j})^{k+1}:\label{HintCOrb}
\end{equation}
where the first sum runs over all the sites whereas the second sum runs over the different orbitals on the same site. Following Ref.~\cite{regnault-PhysRevX.1.021014}, we remove the spurious  effect of band dispersion and band mixing by using only the Hilbert space of the $i$-th band and neglecting its kinetic energy (standard flat band procedure). We assume that we have infinite band gaps and the $i-1$ first bands are filled and inert. The filling factor $\nu$ is defined with respect to the partially filled band, which has Chern number $C$. 

\begin{center}
\begin{figure}[htb]
\includegraphics[width= \linewidth]{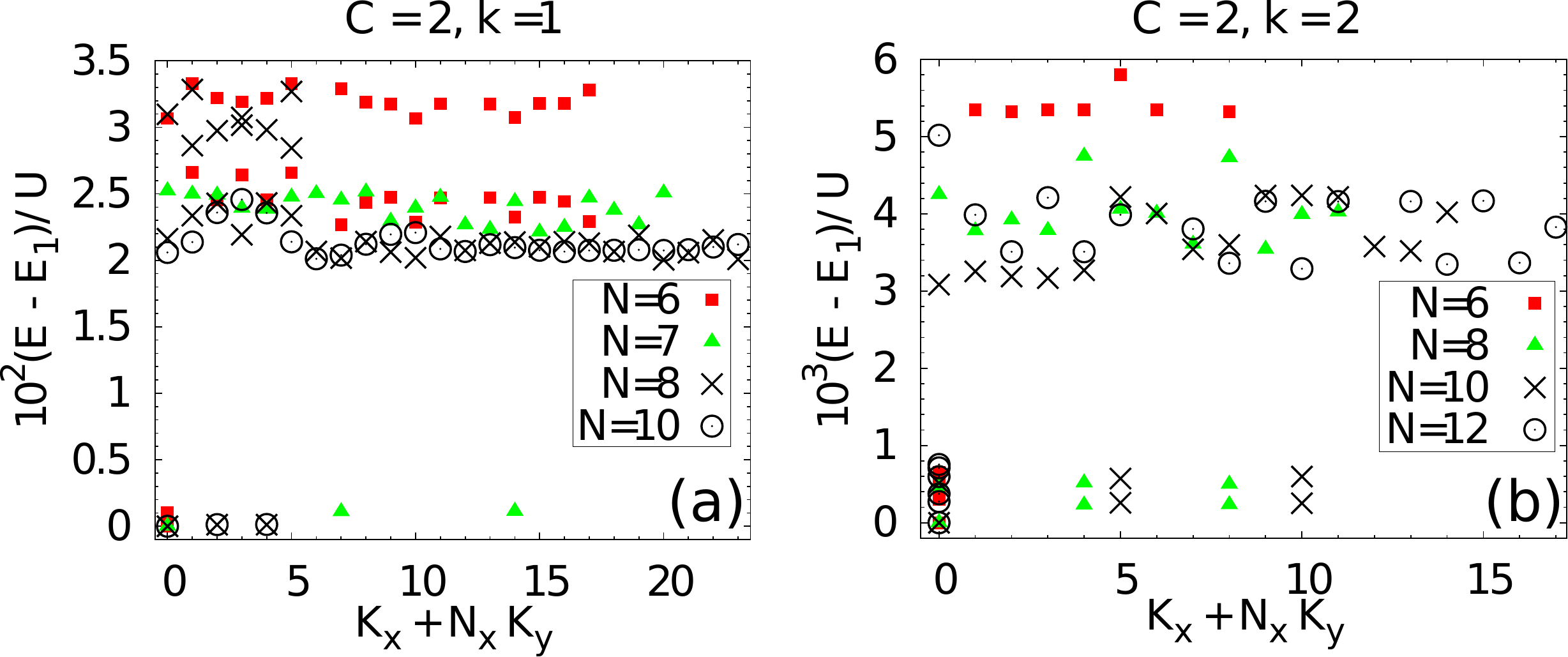}
\caption{(a): Low energy spectra on the pyrochlore lattice with $C=2$ and two body interaction for $N_e = 6, 7, 8, 10$ bosons at $\nu = \frac{1}{3}$. (b): Low energy spectra on the pyrochlore lattice with $C=2$ and three body interaction for $N_e = 6, 8, 10, 12$ bosons at $\nu = 2/3$, we observe an almost sixfold degenerate groundstate only for an even number of particles.}
\label{ground_state_pyrochlore_c_2}
\end{figure}
\end{center}

As very recently reported in Ref.~\onlinecite{Wang-PhysRevB.86.201101} for Chern number $C = 2$ and in Ref.~\onlinecite{Liu-PhysRevLett.109.186805} for $C = N$, there is clear evidence of Abelian phases at filling $\nu = 1/(C+1)$ for bosons with $H_{{\rm int}, 1}$. The energy spectra for $C = 2,3$ are shown in Fig.~\ref{ground_state_pyrochlore_c_2}a)  and  Fig.~\ref{ground_state_pyrochlore_c_3}a) respectively. For both cases and for all numerically accessible numbers of particles, we find $(C+1)$-fold quasi-degenerate groundstates. One can generate quasiholes by adding unit cells to the system. While performing this operation, we find that the \emph{total} quasihole states counting also matches the total number of partitions obeying the $(1,C+1)_1$ generalized Pauli principle, suggesting an Abelian state.

\begin{widetext}
\begin{center}
\begin{figure}[htb]
\includegraphics[width= 0.8\linewidth]{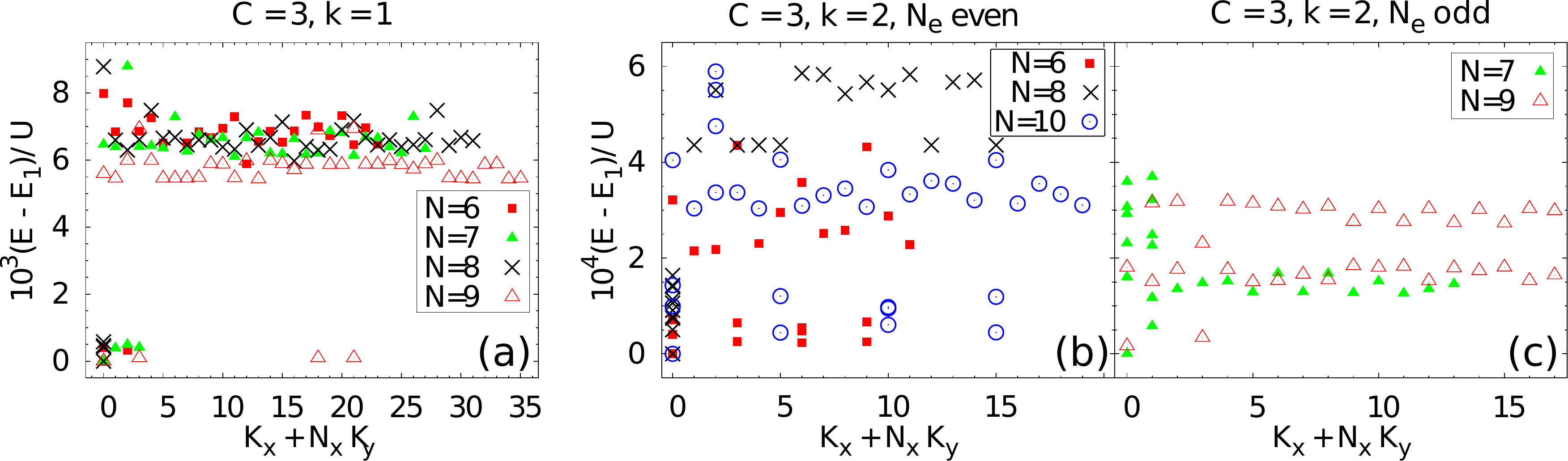}
\caption{(a): Low energy spectra on the pyrochlore lattice with $C=3$ and two body interaction for $N_e = 6, 7, 8, 9$ bosons at $\nu = \frac{1}{4}$. (b): Low energy spectra on the pyrochlore lattice with $C=3$ and three body interaction for $N_e = 6, 8, 10$ bosons at $\nu = 1/2$. (c): Low energy spectra on the pyrochlore lattice with $C=3$ and three body interaction for $N_e = 7, 9$ bosons at $\nu = 1/2$. We observe an almost tenfold degenerate groundstate only for even number of particles. We only show the lowest energy per momentum sectors in addition to the degenerate groundstate. The energies are shifted by $E_1$, the lowest energy for each system size. For $C=3$ three body, only the $N_e$ even sector exhibits a full gap between the degenerate groundstate manifold and the excited states, a signature of pairing. }
\label{ground_state_pyrochlore_c_3}
\end{figure}
\end{center}
\end{widetext}

\begin{figure}[htb]
\includegraphics[width=0.98\linewidth]{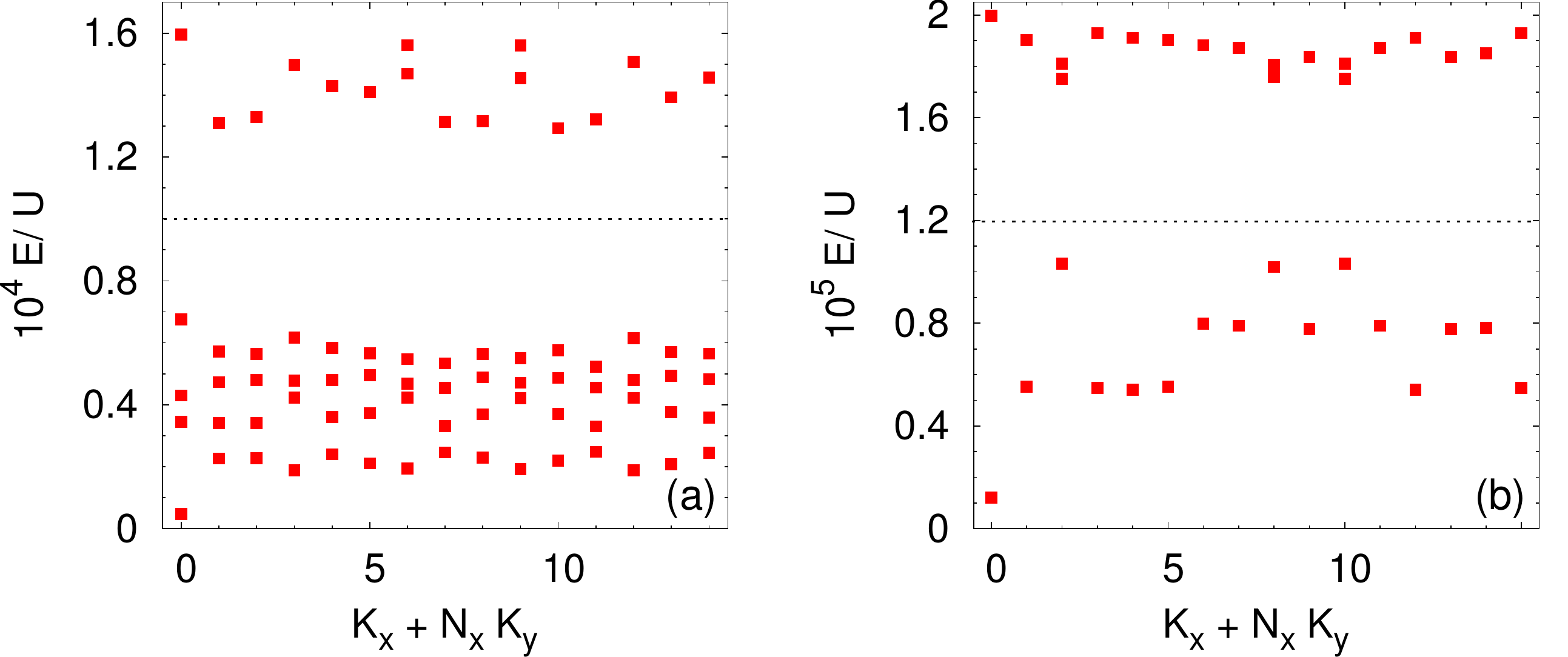}
\caption{Low energy spectra for the $N_e = 7$ bosons on a $(N_x, N_y)=(3,5)$ pyrochlore lattice with $C=3$ and three-body interaction (one site added compared to the $\nu = 2/4$ groundstate) (a) and for the $N_e = 7$ bosons on a $(N_x, N_y)=(4,4)$ pyrochlore lattice (two sites added compared to the $\nu = 2/4$ groundstate) (b). The number of states below the gap (materialized by a dashed line) --- respectively $60$ (a) and $480$ (b) --- is in agreement with the $(2,4)_1$ counting.}
\label{QH_pyrochlore_c_3_k_2}
\end{figure}

Using the three-body interaction ($k=2$), we find a six-fold quasi-degenerate groundstate at filling $\nu = 2/3$ for $C=2$ and a ten-fold quasi-degenerate groundstate at filling $\nu = 1/2$ for $C=3$ and even numbers of particles. The energy spectrum for $C=2$ is shown on Fig.~\ref{ground_state_pyrochlore_c_2}b) while the one for $C = 3$ is shown on Fig.~\ref{ground_state_pyrochlore_c_3}b). The \emph{total} number of groundstates and quasihole states (depending on the system size, from 1 to 10 added unit cells) are compatible with the $(2, C +1)_1$ generalized Pauli principle \cite{bernevig-PhysRevLett.100.246802}, which suggests a paired, non-Abelian state. Figs.~\ref{QH_pyrochlore_c_3_k_2}a) and~\ref{QH_pyrochlore_c_3_k_2}b) show the energy spectrum for one and two added quasiholes in the $C=3$ case. Another signature of pairing comes from the even-odd particle number aliasing for $C=3$. While for $C=2$ the groundstate can only be realized with $N_e$ even due to the filling factor,  this constraint does 
not apply for $C=3$. In that case, we only observe a gap separating the almost tenfold degenerate groundstate manifold from the higher energy excitation for $N_e$ even (see Figs.~\ref{ground_state_pyrochlore_c_3}b) and \ref{ground_state_pyrochlore_c_3}c), a clear signature of pairing.   

For a general $(k+1)$-body interaction, we expect to find a topological phase at filling $\nu = \frac{k}{C+1}$, for which the groundstates and quasiholes states countings are given by the $(k, C +1)_1$ generalized Pauli principle. However, in every model investigated so far, we have only found weak or conflicting evidence (such as correct groundstate degeneracy but no entanglement gap) of these phases for $k>2$. One should keep in mind that FCI states are highly dependent on the underlying tight binding model and interaction range. Thus the absence of the full series is most probably related to the peculiarities of the models used than due to a fundamental reason. Among the other models we have studied, we also observe clear signatures for the following states: $k=2,C=2$ in the triangular lattice model\cite{Wang-PhysRevB.86.201101}, $k=1,C=3$ in the two orbitals on a triangular lattice\cite{Yang-PhysRevB.86.241112} and $k=1,C=4,5$ in the $C$ orbitals on a square lattice model\cite{Yang-PhysRevB.86.241112}.

For the cases studied, while the total number of groundstates and quasiholes states is compatible with a $(k, C +1)_1$ generalized Pauli principle, differences appear in the observed counting per momenta sector: it does not match the counting obtained using the FQH to FCI mapping\cite{Bernevig-2012PhysRevB.85.075128} developed for $C=1$ systems. In particular, when the Pauli principle applied is an inherently ``fermionic'' one on the FQH side (as $(1,3)_1$, the one for the fermionic Laughlin state at $\nu = 1/3$), the same Pauli principle applied for bosons gives a counting in the reduced Brillouin zone which does not have the expected $\mathcal{C}_4$ symmetry. This shows the need for the development of a new folding mapping for the $C>1$ systems \cite{Yangle-WuFuturePaper}.

While the total counting of degenerate groundstates and multiplets matches a spinless FQH counting, this is \emph{not} evidence that the state does not have some internal $SU(C)$ symmetry. One can analytically show that for a fixed value of $N_e$ and $N_\Phi$, each $(k,r)_C$ spinful admissible partition can be mapped onto a spinless $(k,C(r - 1) + 1)_1$ admissible partition with the same number of particles and $C N_\Phi$ orbitals. If $\{(\lambda_i, \sigma_i)\}$ is a $(k,r)_C$ $SU(C)$ admissible partition, then the corresponding spinless partition is $\{\tilde{\lambda}_i=C \lambda_i + \sigma_i\}$. In our case, $r=2$. Using the relation $\nu_{\mathrm{FQH}}=C \nu$, we obtain that $N_x N_y=C N_\Phi$. It is then clear that the counting of groundstates and excitations in the energy spectrum alone cannot differentiate between spinless and spinful (color) FCI states. In particular, a feature of these FCI models at $k=1$ has not been pointed out in previous studies: a $(C+1)$-fold almost degenerate groundstate manifold appears at $\nu=1/(C+1)$ for every $N_x N_y$ values while in the usual spinful FQH (without dislocation\cite{Barkeshli-PhysRevB.81.045323}) one expects to observe them only when this number is a multiple of $C$. This property, which also holds true for quasiholes states and any $k$, would argue for a spinless states. Additional numerical results are available in the supplementary material \cite{SuppMeta}.

\section{Entanglement spectroscopy}\label{sec:pes}

More information about the nature of the groundstate can be obtained from the particle entanglement spectrum\cite{li-08prl010504,sterdyniak-PhysRevLett.106.100405} (PES). For a $d$-fold degenerate state $\{|\psi_i>\}$, we consider the density matrix $\rho=\frac{1}{d}\sum_{i=1}^{d}|\psi_i><\psi_i|$. We divide the $N_e$ particles into two groups $A$ and $B$ with respectively $N_A$ and $N_B$ particles. Tracing out on the particles that belong to $B$, we compute the reduced density matrix $\rho_A={\rm Tr}_B \rho$. The symmetries of the original state, preserved by the operation, allow to label the eigenvalues $\exp(-\xi)$ of $\rho_A$ by their corresponding quantum numbers. For example in the case of a spinful $SU(C)$ eigenstate, both the eigenvalues of the Cartan subalgebra and Casimir operators and the momentum could be used to label the PES eigenvalues\cite{ardonne-PhysRevB.84.205134}. For spinless FQH model states, the number of non zero eigenvalues in $\rho_A$ \emph{exactly matches} the number of quasiholes states for $N_A \le N/2$ particles and the same number of flux quanta as the original state. The quasiholes states counting is characteristic of each topological spinless state and thus the PES acts as a fingerprint of the phase (obtained only from the groundstate), also able to differentiate it from a CDW, which the energy spectrum is not\cite{Bernevig-2012arXiv1204.5682B}. In FCI, since the groundstates deviate from model wavefunctions, one expects to observe a low energy structure similar to the one of the model state with a gap to higher energy excitations. Such a feature has been shown for Laughlin and MR-like states in FCI \cite{regnault-PhysRevX.1.021014, Wu-2012PhysRevB.85.075116}.

\begin{figure}[htb]
\includegraphics[width=0.88\linewidth]{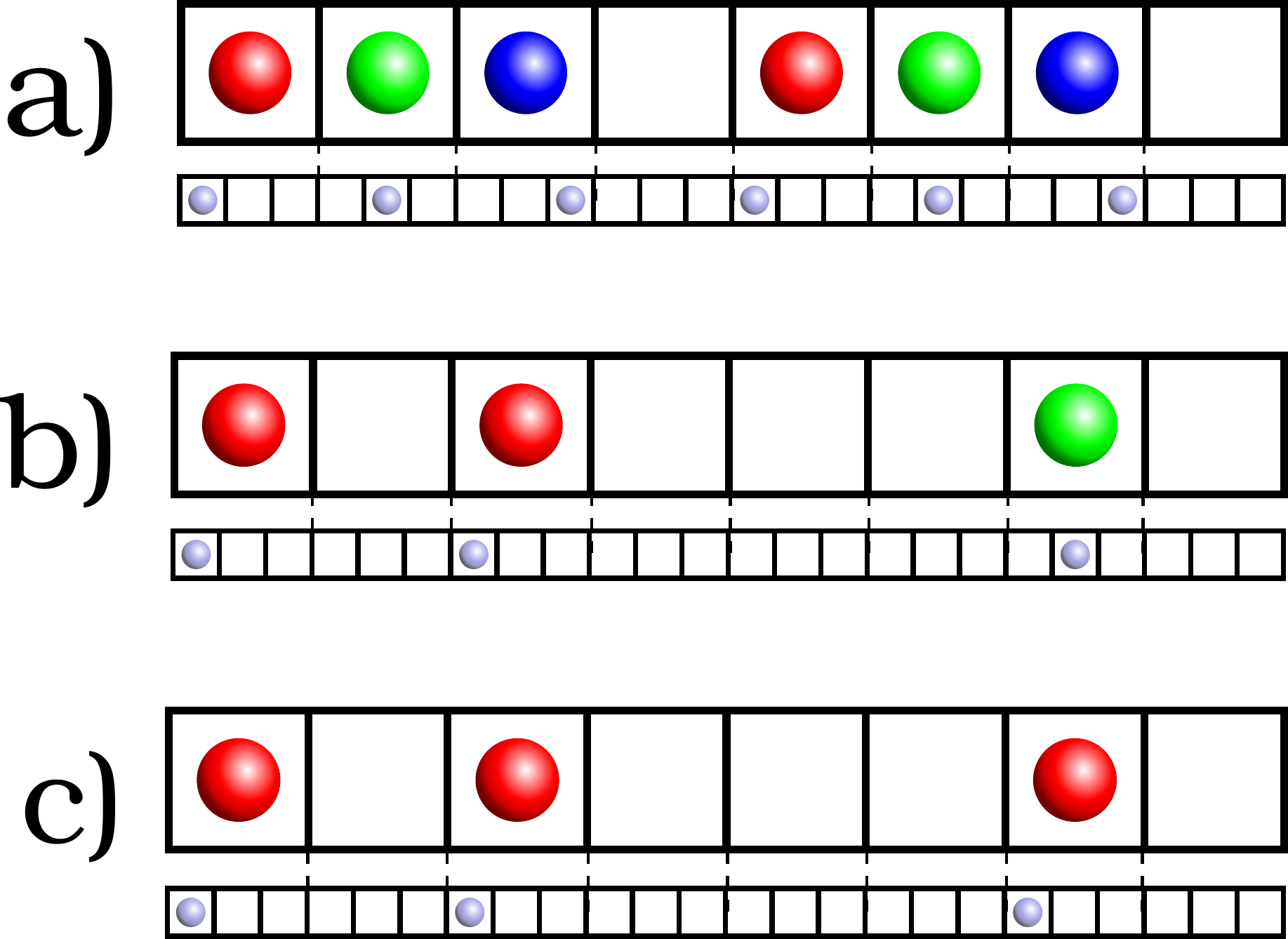}
\caption{$(1,2)_3$ admissible partitions in the $SU(3)$ case (using the red, green and blue colors for each possible value of the internal degree of freedom) and their $(1,4)_1$ counterparts (in gray). Here we have chosen red to be $\sigma =0$, green $\sigma =1$ and blue $\sigma =2$. (a) displays a typical admissible configuration for the groundstate of the $SU(3)$ Halperin state. (b) is an admissible configuration for the quasihole states that is present when computing the PES for the $N_e=6$ particle groundstate. (c) is also an admissible configuration for the quasihole states but cannot be accessed through the PES from the $N_e=6$ particle groundstate.}
\label{partitionsu3}
\end{figure}

When the  groundstate possesses an additional symmetry preserved by particle partitioning, like the $SU(C)$ symmetry for Halperin state in Eq.~\ref{halperin},  an additional constraint can reduce the number of eigenvalues of the PES from the number of quasiholes states of $N_A$ particles in the original number of fluxes. For instance, Halperin groundstates are spin singlets and hence the number of particles per color in them is the same and equal to $N_e/C$. If $N_A> N_e/C$, the quasihole states with $N_A$ particles in a given color should not and cannot be found in the FQH  PES (as depicted in Fig.~\ref{partitionsu3}). In that case, these states and their full corresponding $SU(C)$ multiplet have to be removed to obtain the PES counting. From this perspective, the $C=3$ is a perfect test case since $C=2$ does not provide any additional constraint compared to the spinless case due to the constraint on the entanglement spectrum $N_A \le N_e/2$. Indeed, in all the $C=2$ cases we have checked, the PES counting below the entanglement gap matches the spinful $(k,2)_2$ quasihole states counting which is identical to the spinless $(k,2+k)_1$ quasihole states counting. 

\begin{figure}[htb]
\includegraphics[width=0.98\linewidth]{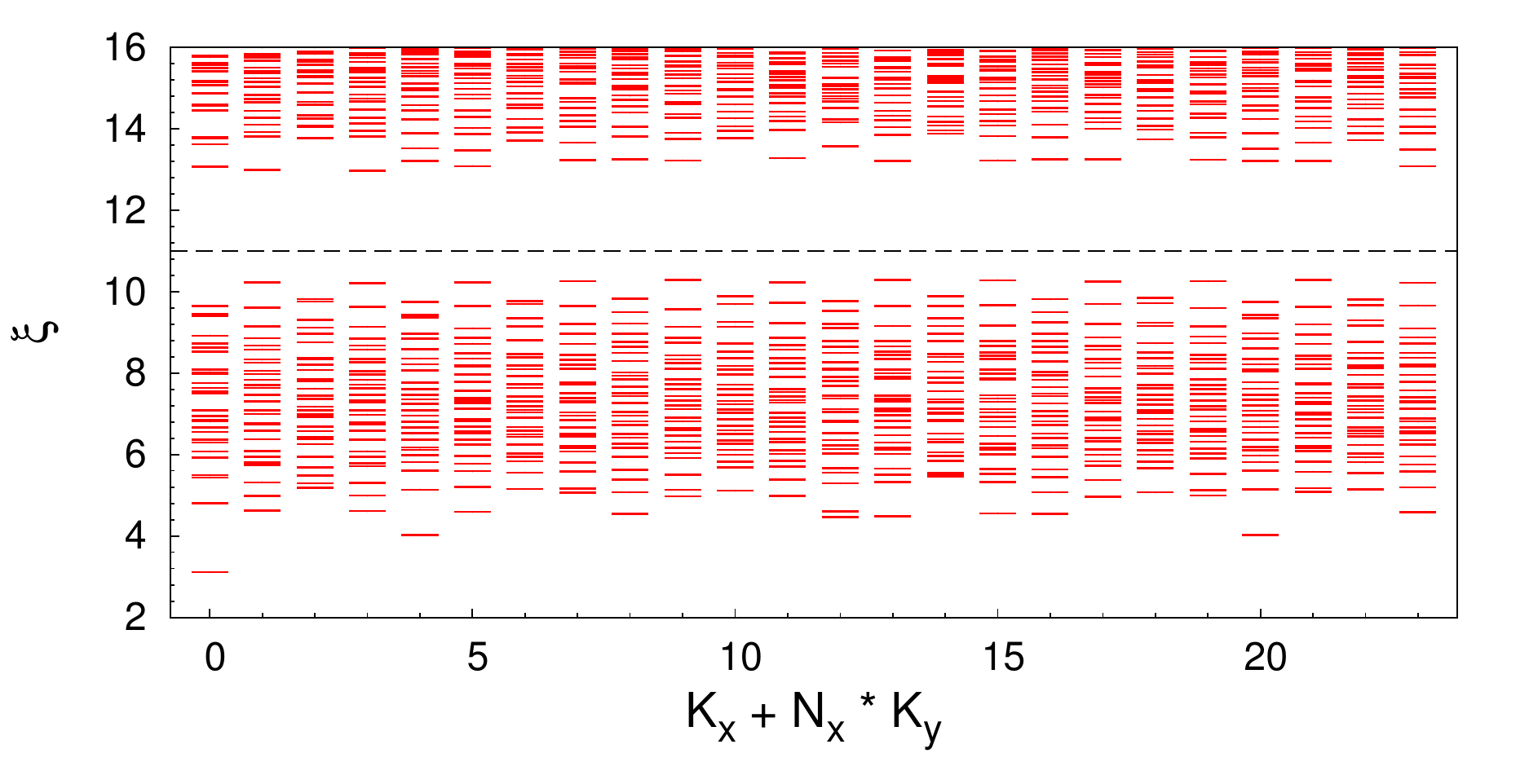}
\caption{PES for low energy groundstate manifold on the pyrochlore lattice with $C=3$ and two body interaction for $N_e = 6$ bosons and $N_A=3$. The number of states below the dotted line is $680$. The $48$ missing states from the $(1,2)_3$ counting are the quasihole states that require 3 particles with the same value of $\sigma$. This counting does not match the counting from the Halperin PES.}
\label{pes_ground_state_pyrochlore_c_3_k_1_n_6}
\end{figure}

\begin{figure}[htb]
\includegraphics[width=0.98\linewidth]{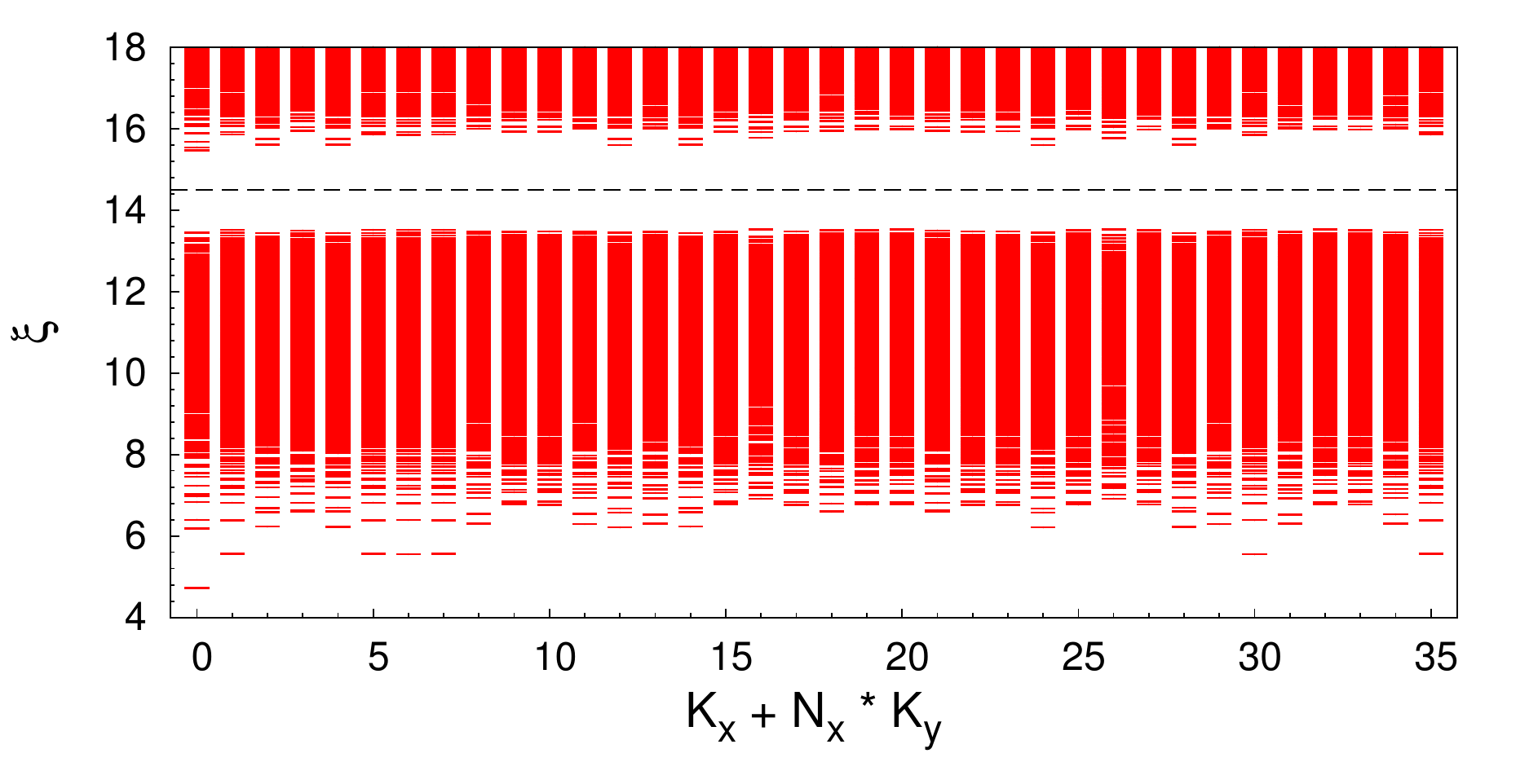}
\caption{PES for low energy groundstate manifold on the pyrochlore lattice with $C=3$ and two body interaction for $N_e = 9$ bosons and $N_A=4$. The number of states below the dotted line is $14364$. This is $1575$ states less than the $(1,4)_1$ counting. This counting matches the counting obtained through the Halperin PES.}
\label{pes_ground_state_pyrochlore_c_3_k_1_n_9}
\end{figure}

We now focus on $C=3,k=1$. Figs.~\ref{pes_ground_state_pyrochlore_c_3_k_1_n_6} and \ref{pes_ground_state_pyrochlore_c_3_k_1_n_9} show the PES for resp. $N_e=6,N_A=3$ and $N_e=9,N_A=4$. These system sizes are directly related to their FQH counterpart \ie $N_x$ or $N_y$ are divisible by $C$. 

In both cases, there is a clear gap with a counting below it that is lower than the spinless $(1,4)_1$ counting: this clearly {\emph{rules out}} a Laughlin-like $\nu=1/4$ state. Only the $N_e=9$ counting matches the Halperin PES counting. The one for $N_e=6$ is higher than expected: it corresponds to the quasihole states counting of the Halperin state where only the states with more than $N_A$ particles of the same color are dropped and not the corresponding full multiplet\cite{Yangle-WuFuturePaper}. When $N_A \le \lceil \frac{N_e}{C} \rceil$, the PES exhibits an entanglement gap above a $(1,4)_1$ counting. 

\section{Conclusion}

We presented strong evidence for the existence of series of Abelian and non-Abelian states at filling $\nu= k/(C+1)$ in fractional Chern insulators with single band of Chern number $C>1$. The energy spectra for both groundstates and quasiholes states suggest that several models support these clustered states. In order to identify the states, we presented evidence from the entanglement spectrum of a hidden $SU(C)$ symmetry of the FCI states that seems to rule out a spinless description. However, we observe several rate anomalies (at particular commensurate particle numbers) in the FCI entanglement spectrum compared to that obtained from the FQH, which could be due to the presence of "effective" twisted boundary conditions in the FCI problem. 

\emph{Acknowledgements}
We thank Y.-L. Wu, B. Estienne, G. M\"oller, Z. Papic and E. Ardonne for useful discussions. AS thanks Princeton University and Microsoft station Q for generous hosting. AS was supported by Keck grant. BAB was supported by Princeton Startup Funds, NSF CAREER DMR-095242, ONR - N00014-11-1-0635, Packard Foundation. NR was supported by  NSF CAREER DMR-095242, ONR - N00014-11-1-0635, Packard Foundation and Keck grant. NR thanks the hospitality of the Aspen Center for Physics supported by the National Science Foundation Grant No. 1066293.

\bibliography{higherchern.bib}

\newpage

\setcounter{section}{0}

\begin{center}
{\bf Supplementary Material to ``Series of Abelian and Non-Abelian States in $C>1$ Fractional Chern Insulators''}
\end{center}

In this Supplementary Material, we provide additional numerical results that might be relevant to a more specialized audience. Our article only showed numerical results for the pyrochlore lattice model~\cite{Trescher-PhysRevB.86.241111}. While this supplementary material provides additional evidences for this system, it also gives numerical results about Abelian and non-Abelian states in $C>1$ fractional Chern insulators  (FCI) for the $C=2$ triangular lattice model~\cite{Wang-PhysRevB.86.201101}, the two-orbitals model on triangular lattice~\cite{Yang-PhysRevB.86.241112} with $C=3$ and the $C$-orbitals model on a square lattice~\cite{Yang-PhysRevB.86.241112}.

\section{pyrochlore}

\begin{figure}[htb]
\includegraphics[width=0.70\linewidth]{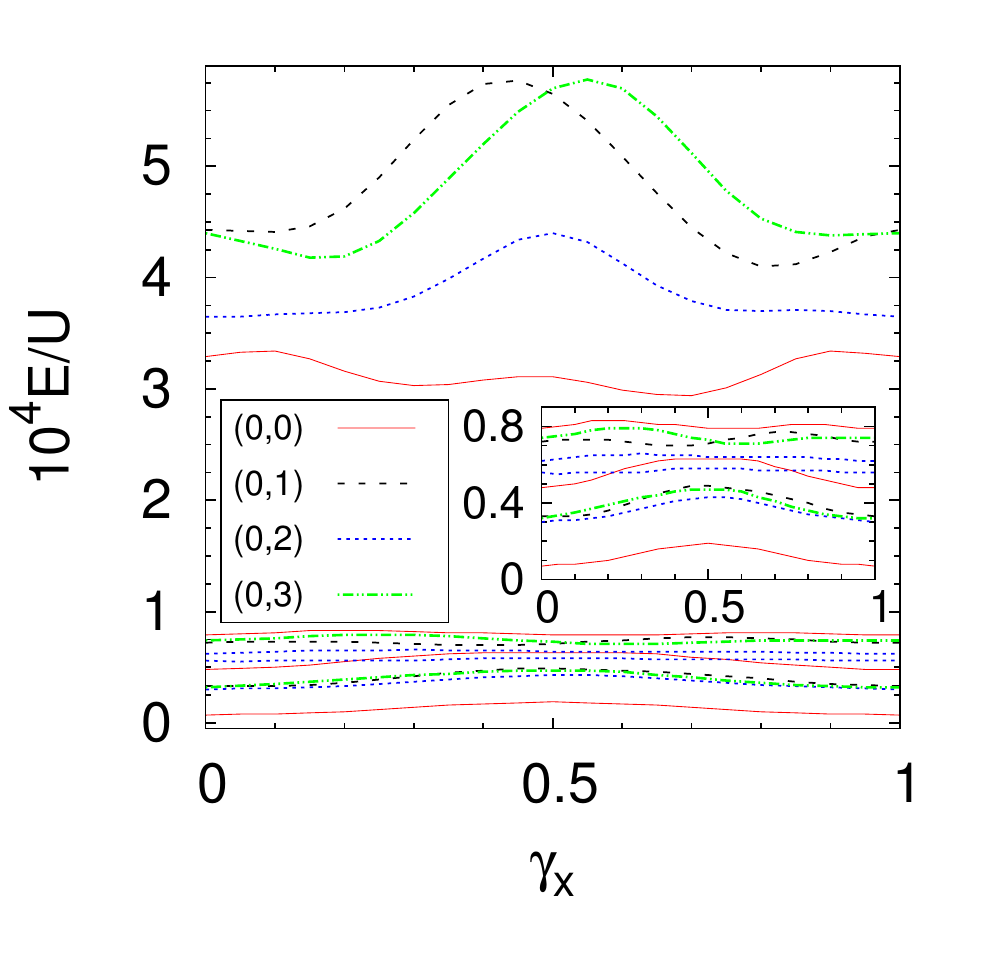}
\caption{Evolution of the low-lying states of the pyrochlore lattice model with $C=3$ and three-body interaction in momentum sectors $(0,0)$, $(0,1)$, $(0,2)$ and $(0,3)$ with $N_e = 6$ bosons on a $(N_x , N_y ) = (3,4)$ lattice upon flux insertion along the $x$ direction. $\gamma_x$ counts the number of inserted flux quanta. We only show the momentum sector $(K_x , K_y ) = (0, 0) , \  (0, 1), \ (0, 2) , \ (0, 3)$  where the almost tenfold degenerate groundstate lies. The inset is a zoom on the low energy part of the spectrum.}
\label{flux_insertion_pyrochlore_c_3_k_2}
\end{figure}

\begin{figure}[htb]
\includegraphics[width=0.80\linewidth]{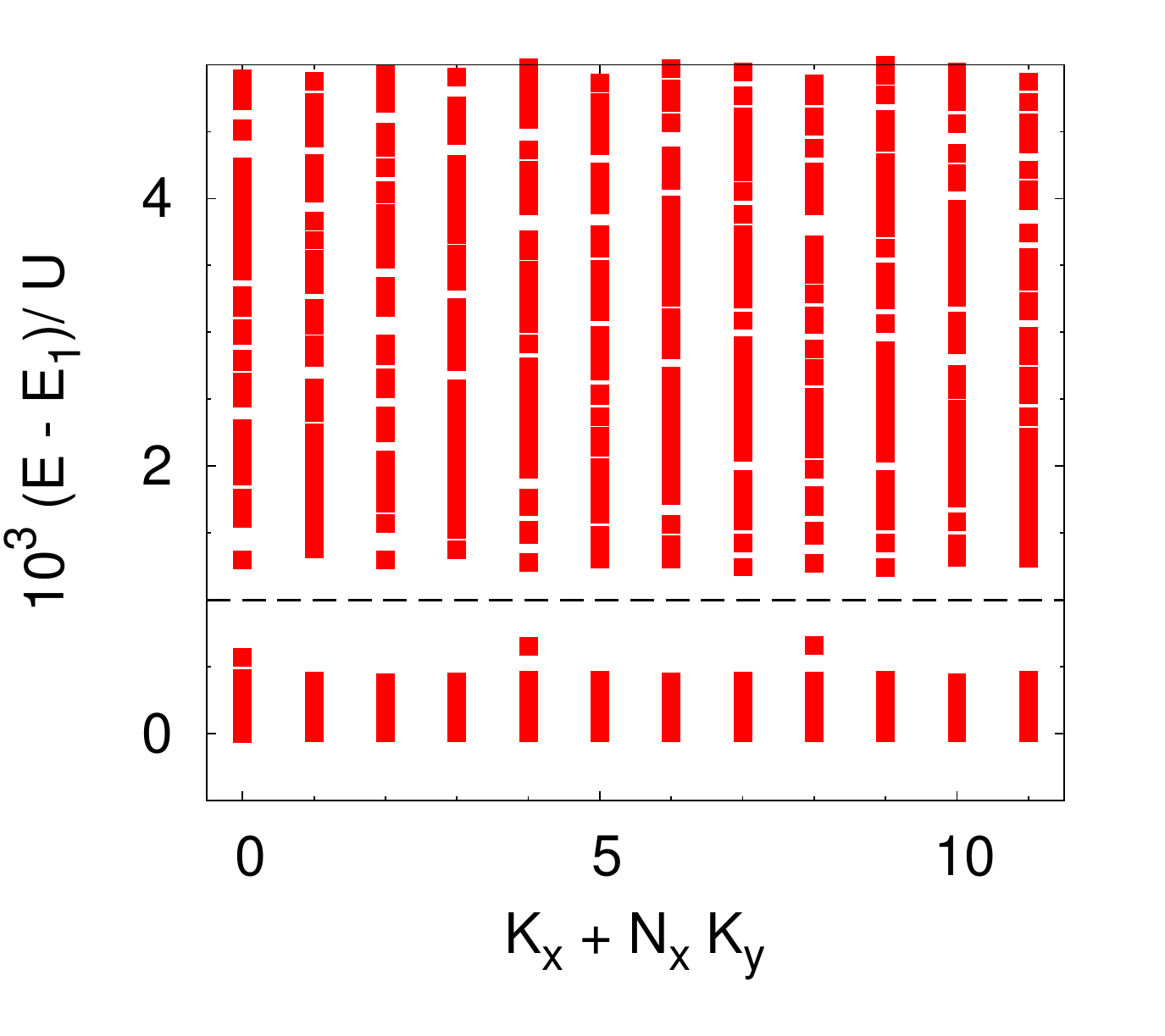}
\caption{Low energy spectra on the pyrochlore $C=3$ model with three-body interaction for the $N_e = 6$ bosons on a $(N_x,N_y)=(4,3)$ pyrochlore lattice (three sites added compared to the $\nu = 2/3$ groundstate). The energies are shifted by $E_1$, the lowest energy for each system size. The number of states below the gap (materialized by a dashed line) --- is equal to $676$ and is in agreement with the $(2,3)_1$ counting.}
\label{QH_pyrochlore_c_2_k_2}
\end{figure}

\begin{figure}[htb]
\includegraphics[width=0.98\linewidth]{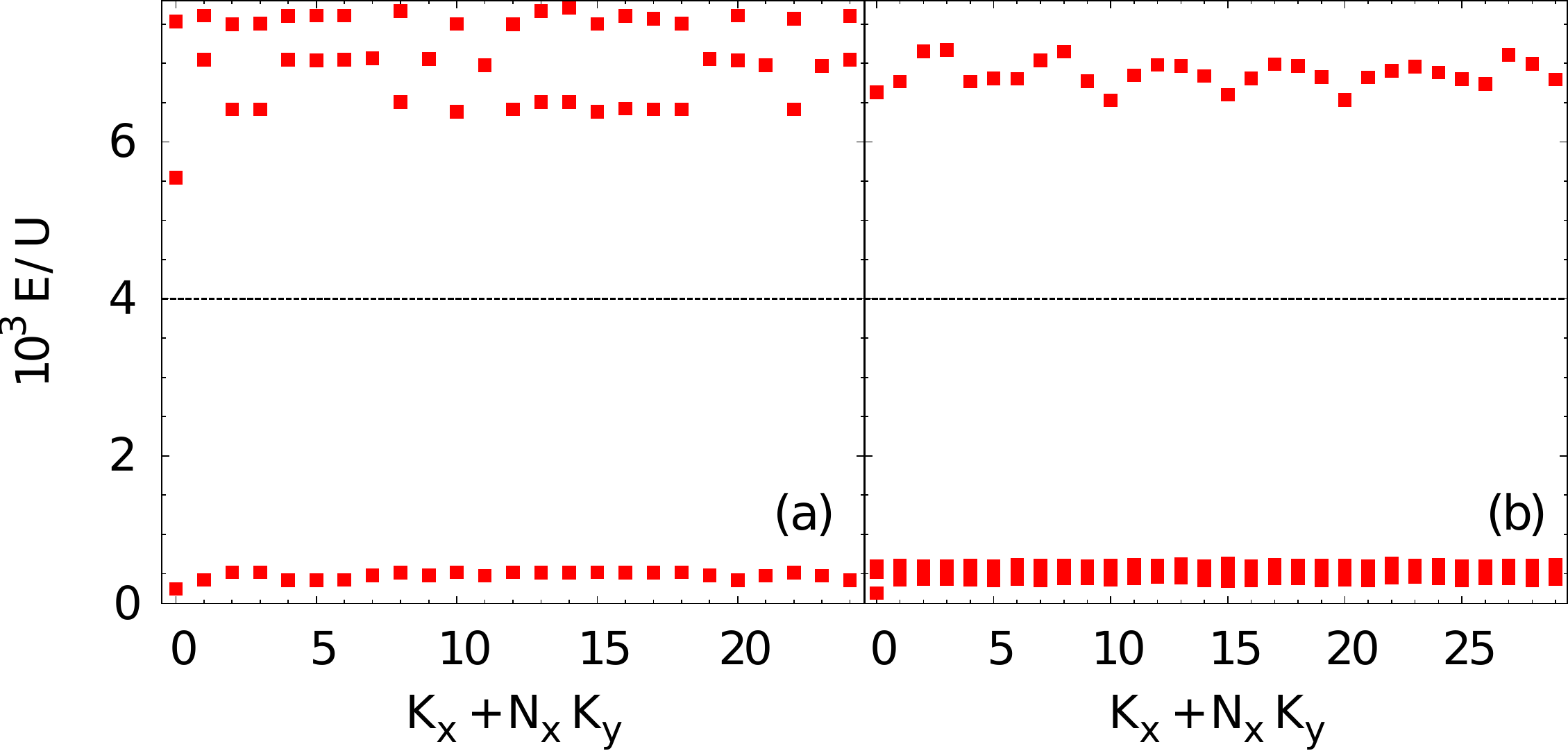}
\caption{Low energy spectra for the $N_e = 6$ bosons on a $(N_x, N_y)=(5,5)$ pyrochlore lattice with $C=3$ and two-body interaction (one site added compared to the $\nu = 1/4$ groundstate) (a) and for the $N_e = 7$ bosons on a $(N_x,N_y)=(5,6)$ pyrochlore lattice (two sites added compared to the $\nu = 1/4$ groundstate) (b). The number of states below the gap (materialized by a dashed line) --- respectively $25$ (a) and $120$ (b) --- is in agreement with the $(1,4)_1$ counting.}
\label{QH_pyrochlore_c_3_k_1}
\end{figure}

In this section, we provide additional evidence for the phase we found on the pyrochlore lattice~\cite{Trescher-PhysRevB.86.241111}. For this model, we have used a $(k+1)$-body Hubbard interaction $H_{{\rm int}, k}=U\sum_{i} :\rho_i^{k+1}:$ where $::$ denotes the normal ordering and the sum runs over all the sites. We checked that upon flux insertion the groundstate manifold does not mix with higher energy states. Also, the insertion of one flux restores the original configuration. This can be observed for $N_e = 6$ particles on a $(N_x , N_y ) = (3,4)$ lattice for $C=3$ with three-body interaction in Fig.~\ref{flux_insertion_pyrochlore_c_3_k_2}. We checked that quasiholes states counting obey the same rules that the groundstate. Some energy spectra for quasiholes are shown on Fig.~\ref{QH_pyrochlore_c_2_k_2} for $C=2$ with three-body interaction, on Fig.~\ref{QH_pyrochlore_c_3_k_1} for $C=3$ with two-body interaction. In both cases, the total observed counting matches the colorful $(k,r)_C$ counting which is identical to the $(k,r+C-1)_1$ counting.

\begin{figure}[htb]
\includegraphics[width=0.98\linewidth]{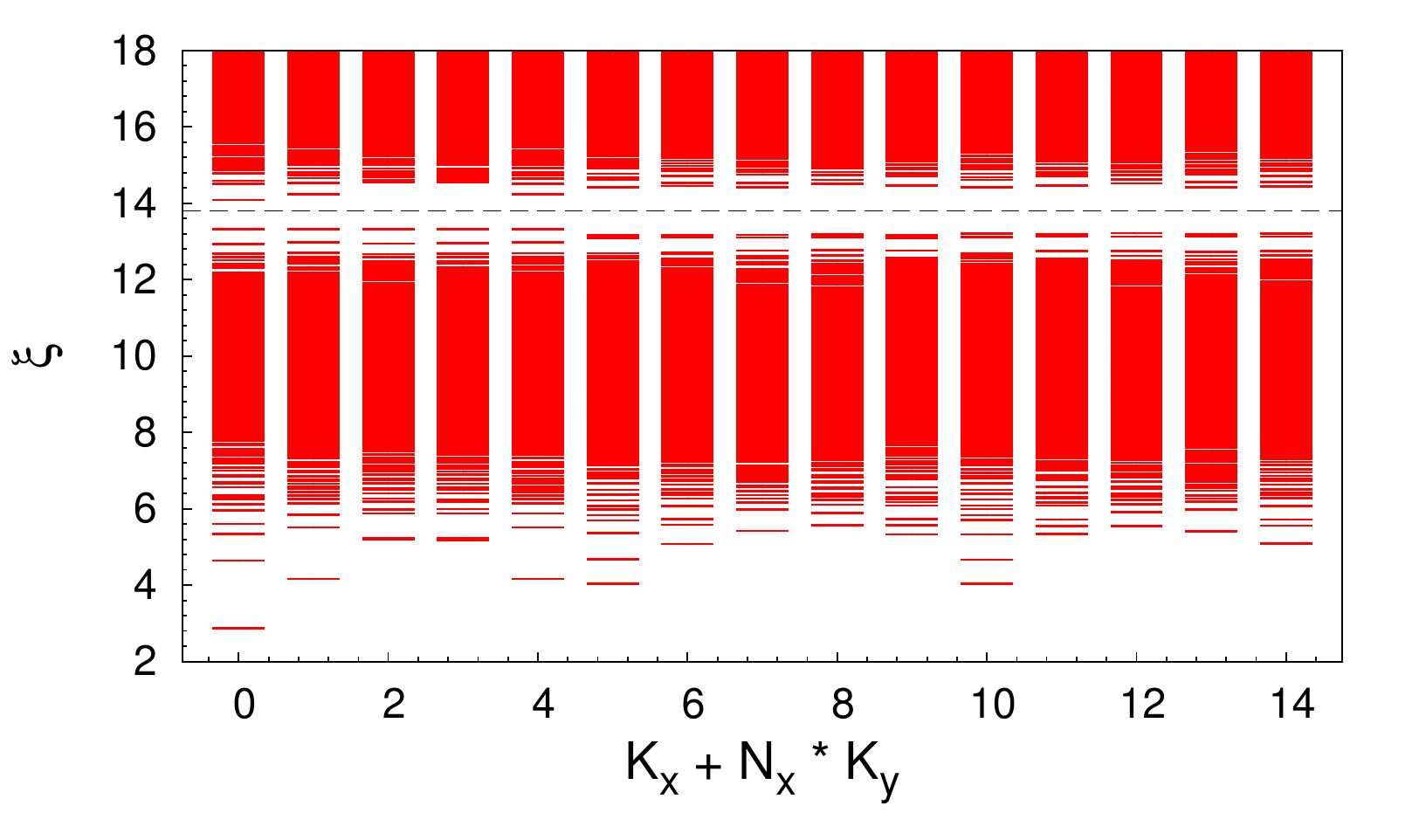}
\caption{PES for low energy groundstate manifold on the pyrochlore lattice with $C=2$ and three-body interaction for $N_e = 10$ bosons and $N_A=5$ on a $(N_x , N_y ) = (5,3)$ lattice. The number of states below the dotted line is $4278$. This is equal to the $(2,3)_1$ counting.}
\label{pes_ground_state_pyrochlore_c_2_k_2}
\end{figure}

\begin{figure}[htb]
\includegraphics[width=0.98\linewidth]{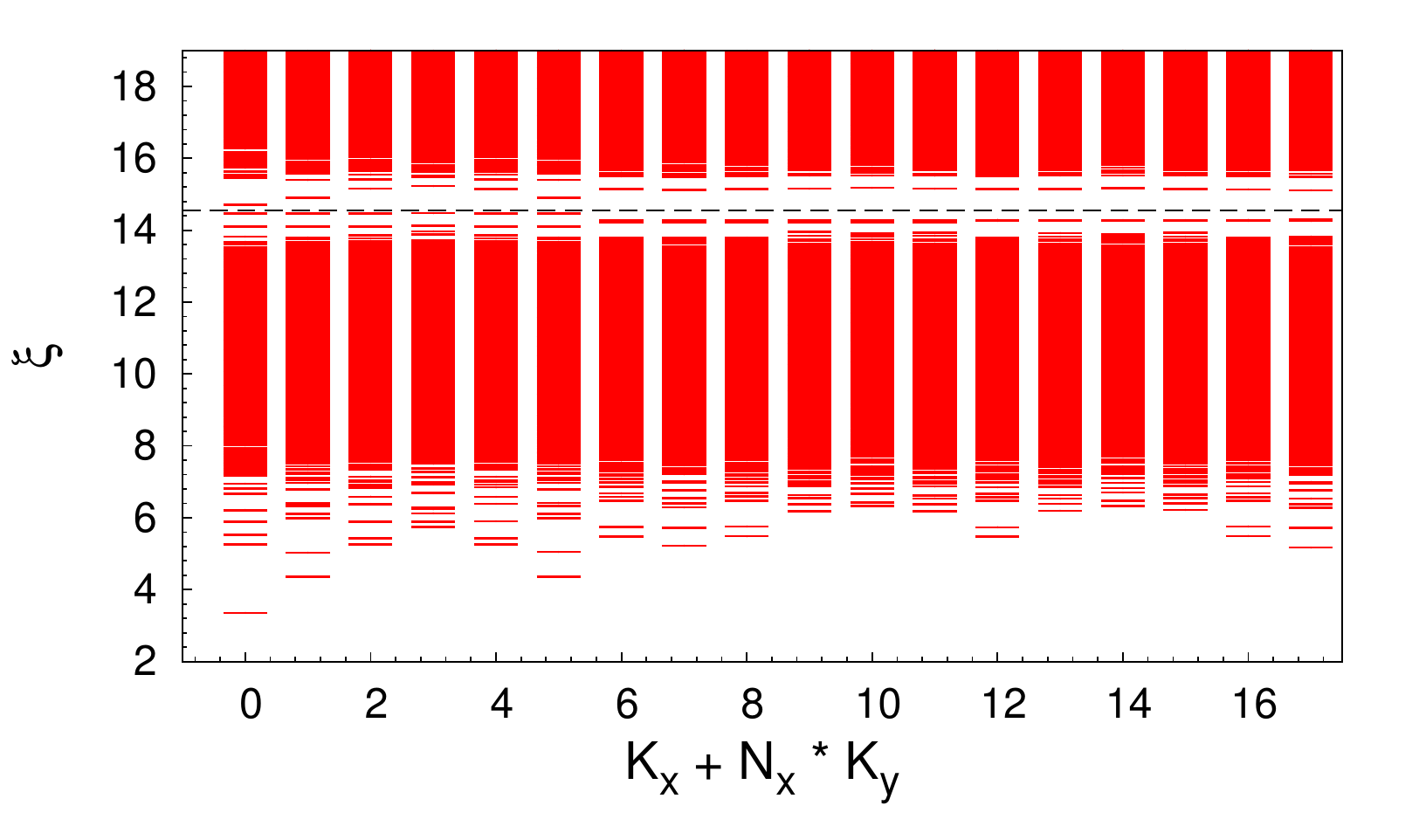}
\caption{PES for low energy groundstate manifold on the pyrochlore lattice with $C=2$ and three-body interaction for $N_e = 12$ bosons and $N_A=5$ on a $(N_x , N_y ) = (6,3)$ lattice. The number of states below the dotted line is $12870$. This is equal to the $(2,3)_1$ counting.}
\label{pes_ground_state_pyrochlore_c_2_k_2_n_12}
\end{figure}

\begin{figure}[htb]
\includegraphics[width=0.98\linewidth]{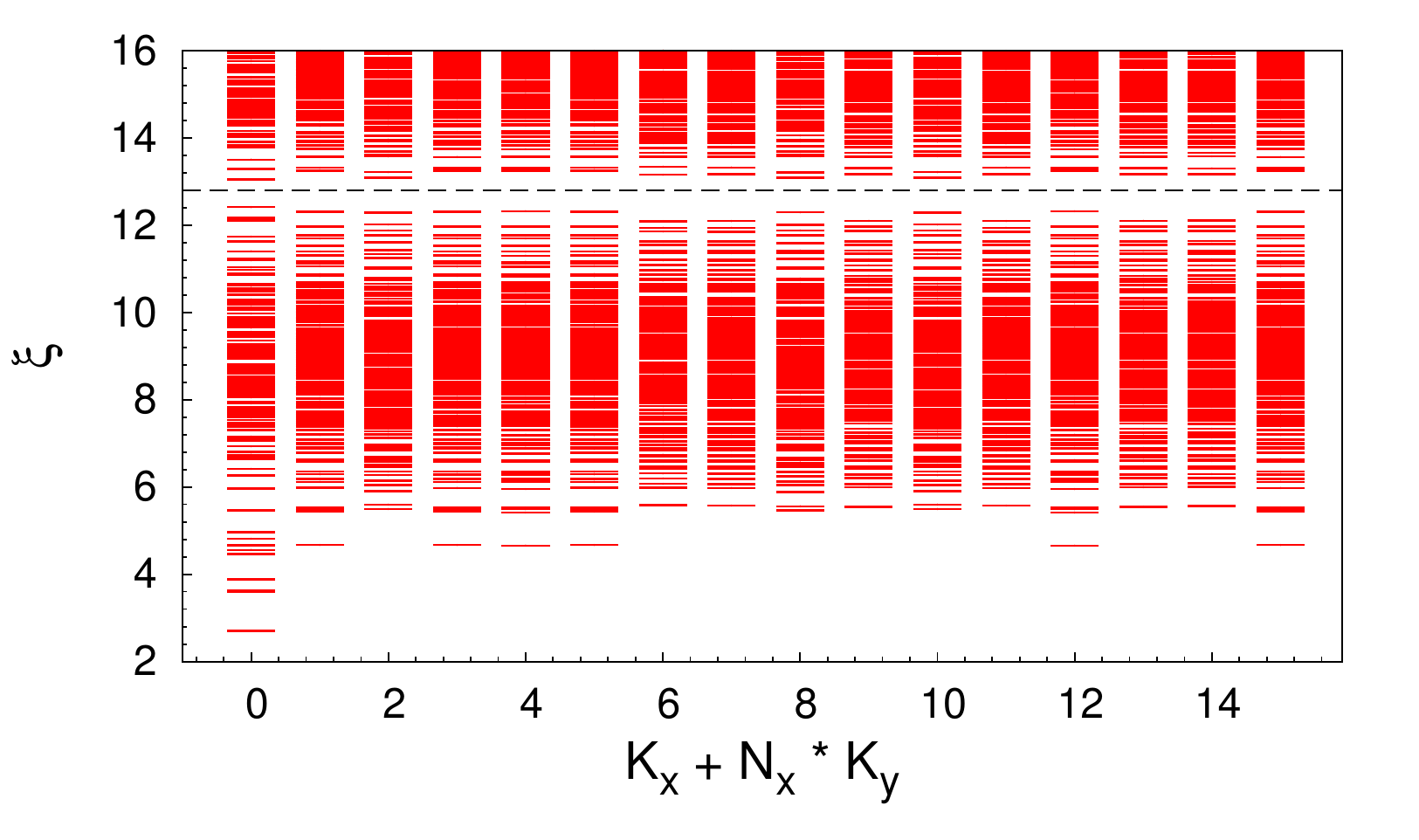}
\caption{PES for low energy groundstate manifold on the pyrochlore lattice with $C=3$ and three-body interaction for $N_e = 8$ bosons and $N_A=4$ on a $(N_x , N_y ) = (4,4)$ lattice. The number of states below the dotted line is $1956$. This is equal to the number of states given by the $(2,2)_3$ counting.}
\label{pes_ground_state_pyrochlore_c_3_k_2}
\end{figure}

As discussed in the article, the counting of the groundstate or quasihole degeneracy does not allow to distinguish between a colorless and colorful physics in the absence of an exact mapping~\cite{Bernevig-2012PhysRevB.85.075128}. But the particle entanglement spectrum (PES)~\cite{sterdyniak-PhysRevLett.106.100405} clearly indicates for $C>2$ that the phase cannot be understood in terms of a colorless phase. We provide additional PES for $C=2$ with three-body interaction on Fig.~\ref{pes_ground_state_pyrochlore_c_2_k_2} and Fig.~\ref{pes_ground_state_pyrochlore_c_2_k_2_n_12}, and for $C=3$ with three-body interaction on Fig.~\ref{pes_ground_state_pyrochlore_c_3_k_2}. As explained on the article, here only the $C=3$ might display a difference from the spinless counting. Note that the case shown on Fig.~\ref{pes_ground_state_pyrochlore_c_3_k_2}, the known counting is still given by the $(2,2)_3$ generalized exclusion principle (which matches the $(2,4)_1$ counting).

\section{$C = 2$ triangular lattice}

\begin{center}
\begin{figure}[htb]
\includegraphics[width= 0.80\linewidth]{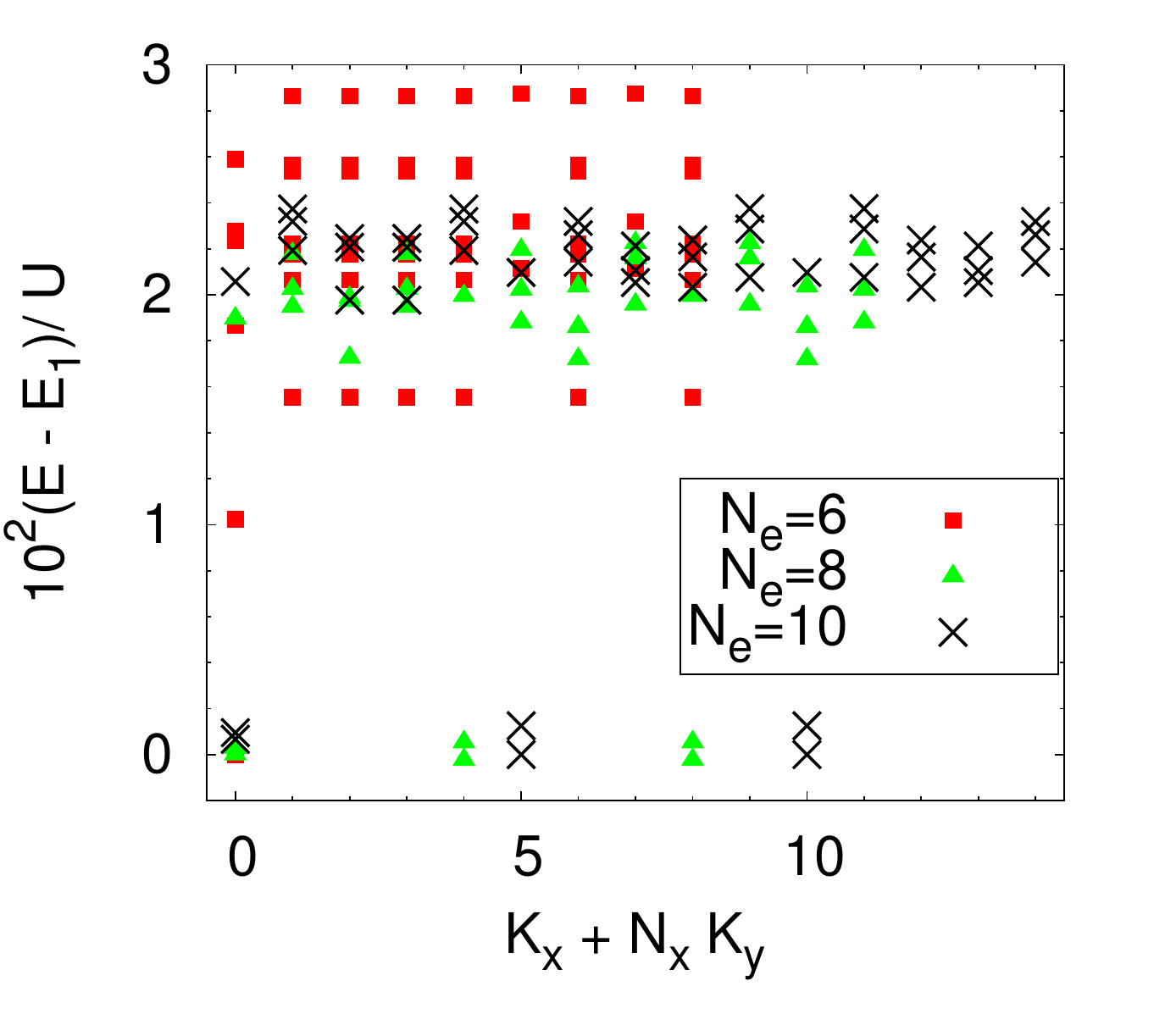}
\caption{Low energy spectra on the triangular lattice with $C=2$ and three-body interaction for $N_e = 6, 8, 10$ bosons at $\nu = \frac{2}{3}$ on a $(N_x , N_y ) = (N_e/2,3)$ lattice. We observe an almost sixfold degenerate groundstate only for an even number of particles.}
\label{ground_state_triangularlattice}
\end{figure}
\end{center}

\begin{figure}[htb]
\includegraphics[width=0.70\linewidth]{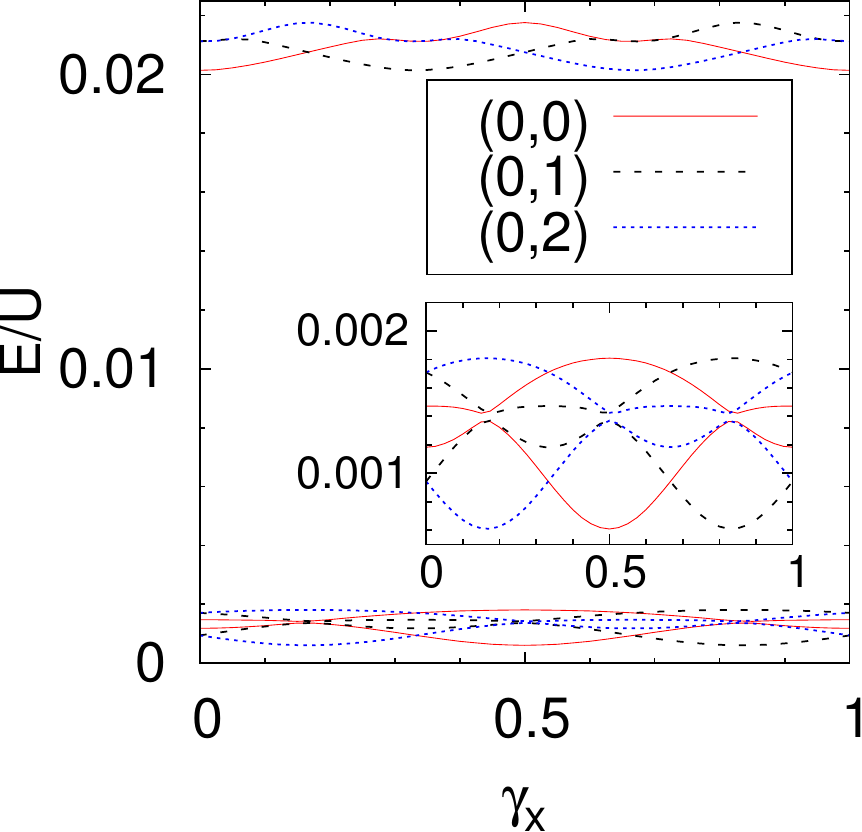}
\caption{Evolution of the low-lying states of the triangular lattice model with $C=2$ and three-body interaction in momentum sectors $(0,0)$, $(0,1)$ and $(0,2)$ with $N_e = 8$ bosons on a $(N_x , N_y ) = (4,3)$ lattice upon flux insertion along the $x$ direction. $\gamma_x$ counts the number of inserted flux quanta. We only show the momentum sector $(K_x , K_y ) = (0, 0) , \  (0, 1), \ (0, 2)$  where the almost sixfold degenerate groundstate lies.}
\label{flux_insertion_triangular_c_2_k_2}
\end{figure}

\begin{figure}[htb]
\includegraphics[width=0.80\linewidth]{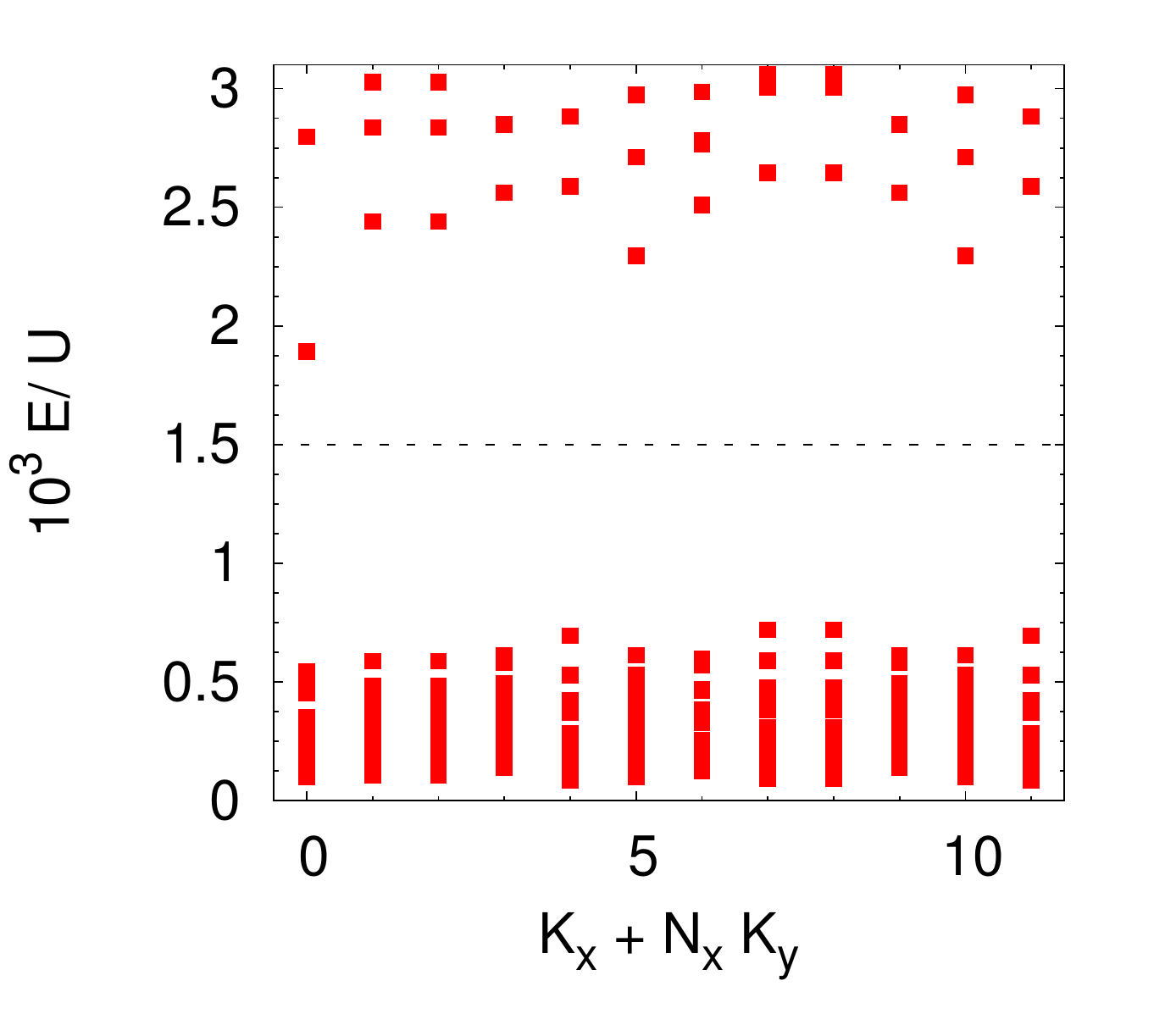}
\caption{Low energy spectra for $N_e = 7$ bosons on a $(N_x , N_y)=(3,4)$ $C=2$ triangular lattice with three-body interaction. The number of states below the gap (materialized by a dashed line) is $144$, in agreement with the $(2,3)_1$ counting.}
\label{QH_triangular_c_2_k_2}
\end{figure}

\begin{figure}[htb]
\includegraphics[width=0.98\linewidth]{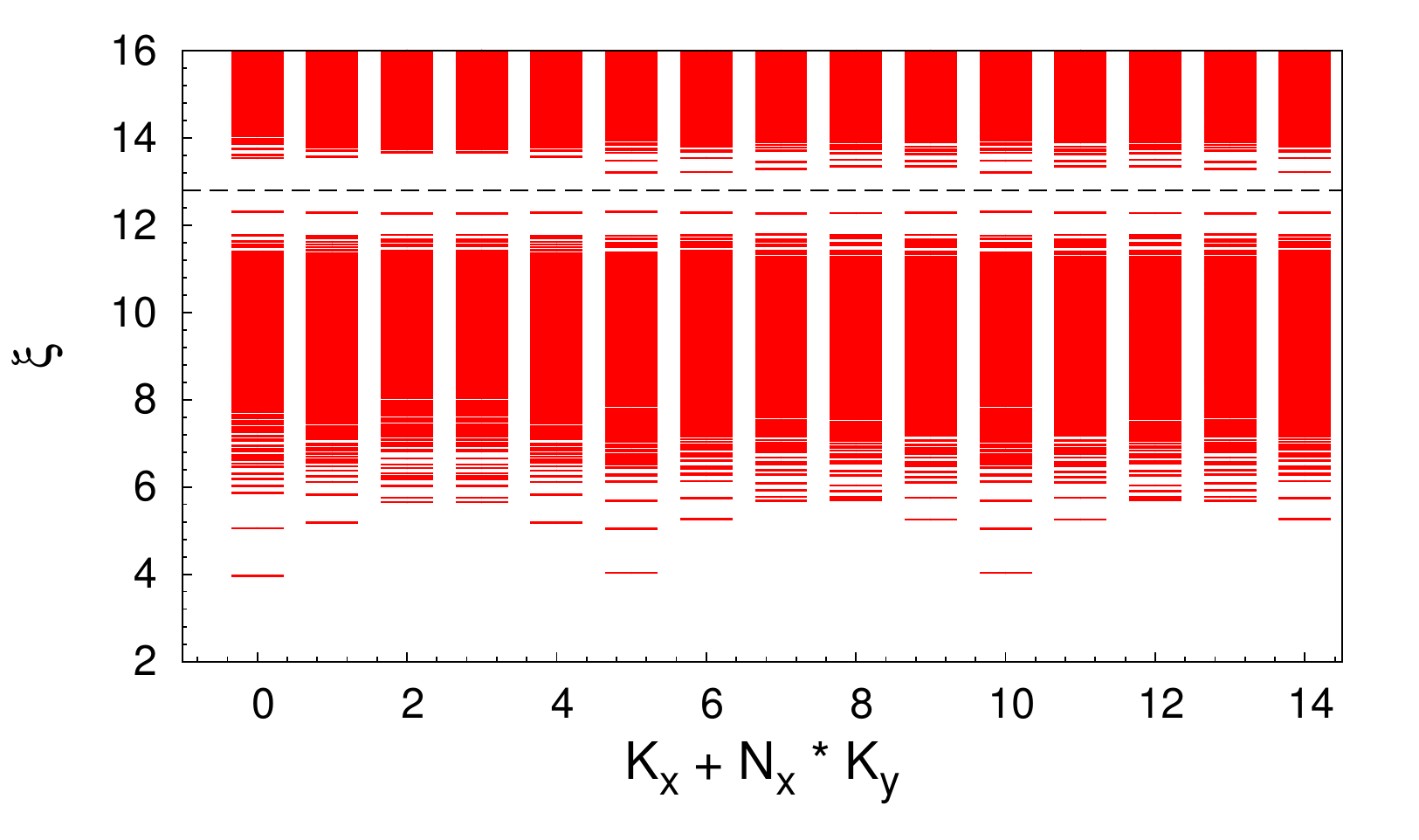}
\caption{PES for low energy groundstate manifold on the triangular lattice with $C=2$ and three-body interaction for $N_e = 10$ bosons and $N_A=5$ on a $(N_x , N_y ) = (5,3)$ lattice. The energies are shifted by $E_1$, the lowest energy for each system size. The number of states below the dotted line is $4278$. This is equal to the $(2,3)_1$ counting.}
\label{pes_ground_state_triangular_c_2_k_2}
\end{figure}

We have investigated the $C=2$ triangular lattice model~\cite{Wang-PhysRevB.86.201101} with three-body interaction, the two-body interaction case have been studied in Ref.~\cite{Wang-PhysRevB.86.201101}. We have used the same parameters as in Ref.~\cite{Wang-PhysRevB.86.201101}. The interaction is given by $H_{{\rm int}, k}=U\sum_{i} :\rho_i^{3} :$. We find convincing evidence of a sixfold degenerate groundstate at $\nu = \frac{2}{3}$ for even numbers of particles. The energy spectra are shown on Fig.~\ref{ground_state_triangularlattice} for several system sizes. We have checked that upon flux insertion the groundstate manifold does not mix with higher energy states. Also, the insertion of one flux restores the original configuration. This can be observed for $N_e = 8$ particles on a $(N_x , N_y ) = (3,4)$ lattice in Fig.~\ref{flux_insertion_triangular_c_2_k_2}. For the quasiholes and the PES, the results are similar to what we have found for the pyrochlore model, as seen respectively on Fig.~\ref{QH_triangular_c_2_k_2} and ~\ref{pes_ground_state_triangular_c_2_k_2}.

\section{$C = 3$ triangular lattice}

\begin{center}
\begin{figure}[htb]
\includegraphics[width= 0.80\linewidth]{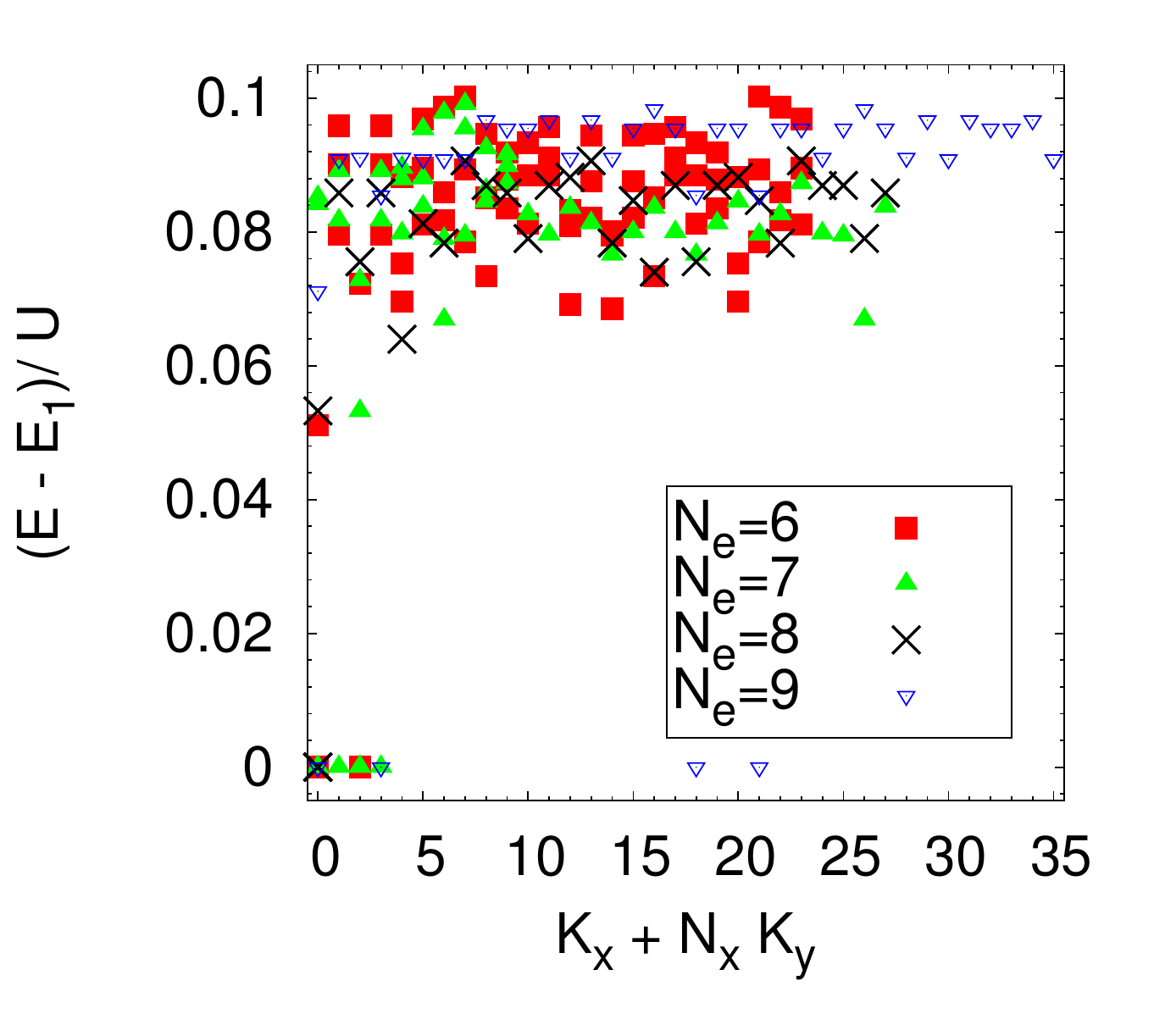}
\caption{Low energy spectra on the $2$-orbitals on triangular lattice with $C=3$ and two-body interaction for $N_e = 6, 7, 8$ bosons at $\nu = \frac{1}{4}$ on a $(N_x , N_y ) = (N_e,4)$ lattice. The energies are shifted by $E_1$, the lowest energy for each system size. As expected, we observe an almost fourfold degenerate groundstate. Note that for $N_e=8$, the four lowest energy states in the $(K_x,K_y)=(0,0)$ are so close in energy that it is impossible to distinguish them.}
\label{ground_state_chern3}
\end{figure}
\end{center}

On the $2$-orbital triangular lattice~\cite{Yang-PhysRevB.86.241112} with $C=3$, we have found a clear signature of a strongly correlated topological phase at $\nu=\frac{1}{4}$ with two-body interaction. We have use the model of ref.~\cite{Yang-PhysRevB.86.241112} with the optimized parameters $t_2/t_1 = 0.28$ and $t_3/t_1 = -0.22$. The interaction is slightly different from the one we have considered previously in order to make contact with the physical system. It consists of an isotropic interaction $H_{{\rm int}, k}=U\sum_{i} :(\sum_j \rho_{i,j})^{2}:$ where the first sum runs over all the sites whereas the second sum runs over the different orbitals on the same site. On Fig.~\ref{ground_state_chern3}, we provide the energy spectrum for the groundstate at $\nu=\frac{1}{4}$ up to $N_e=9$. 

\begin{figure}[htb]
\includegraphics[width=0.80\linewidth]{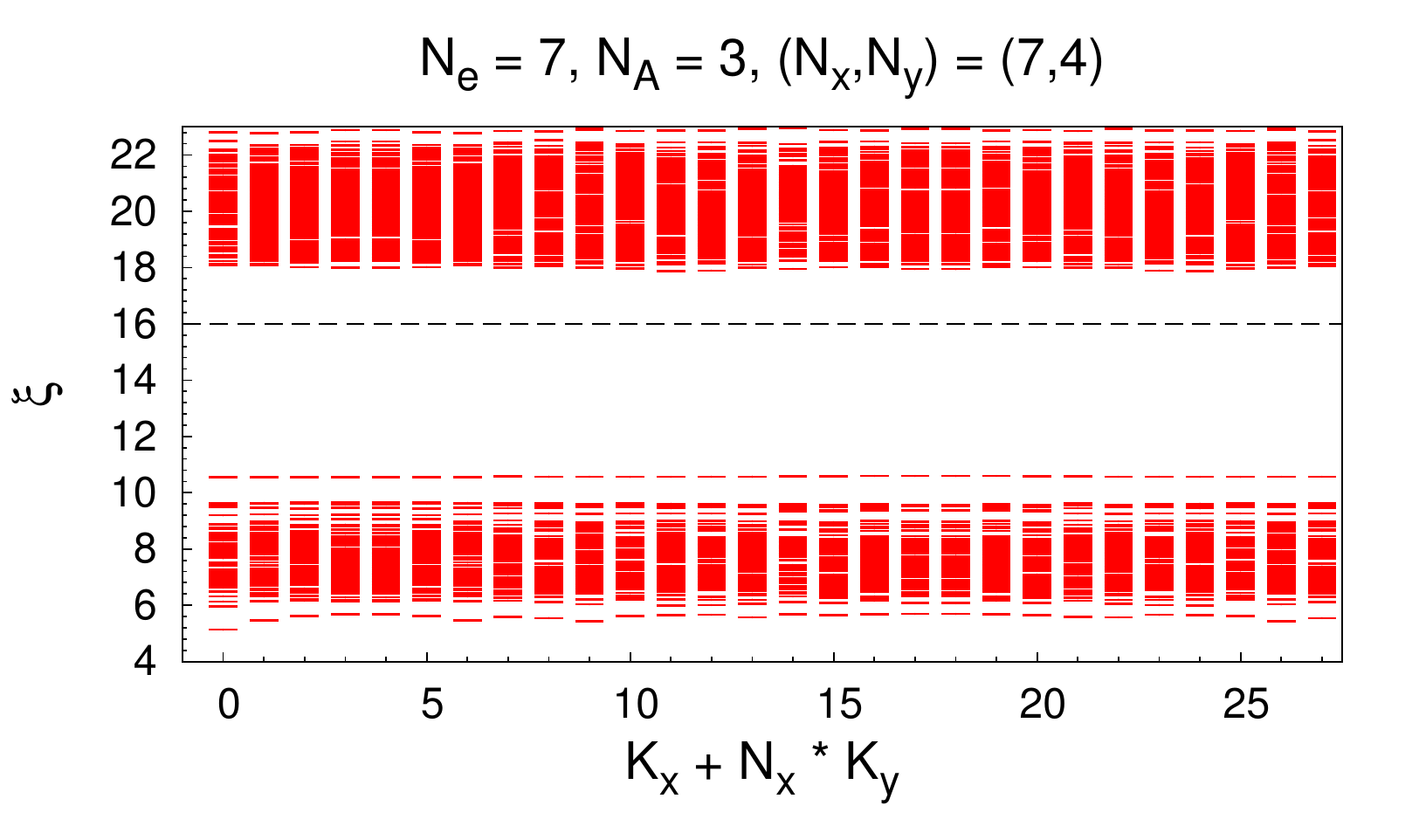}\\
\includegraphics[width=0.80\linewidth]{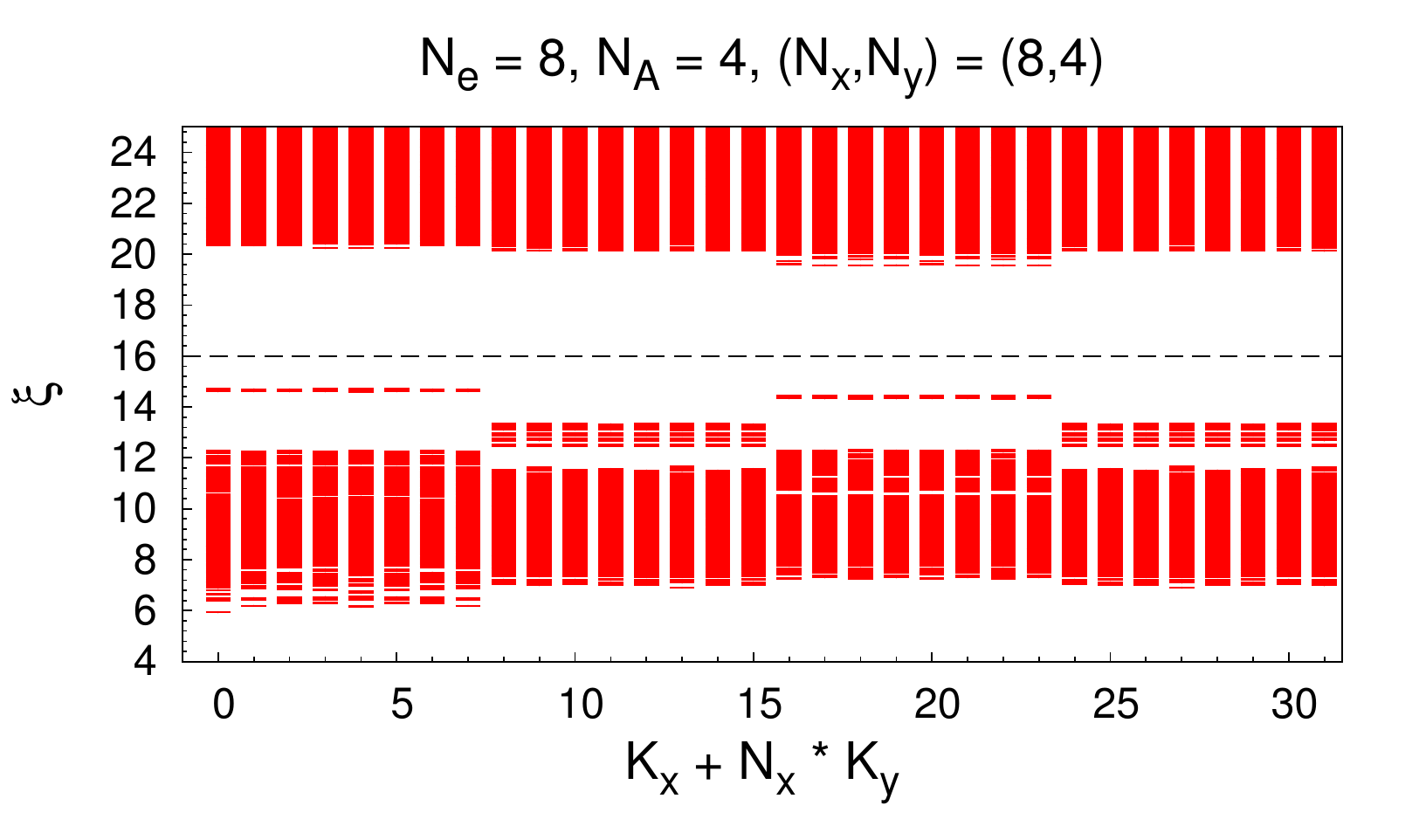}\\
\includegraphics[width=0.80\linewidth]{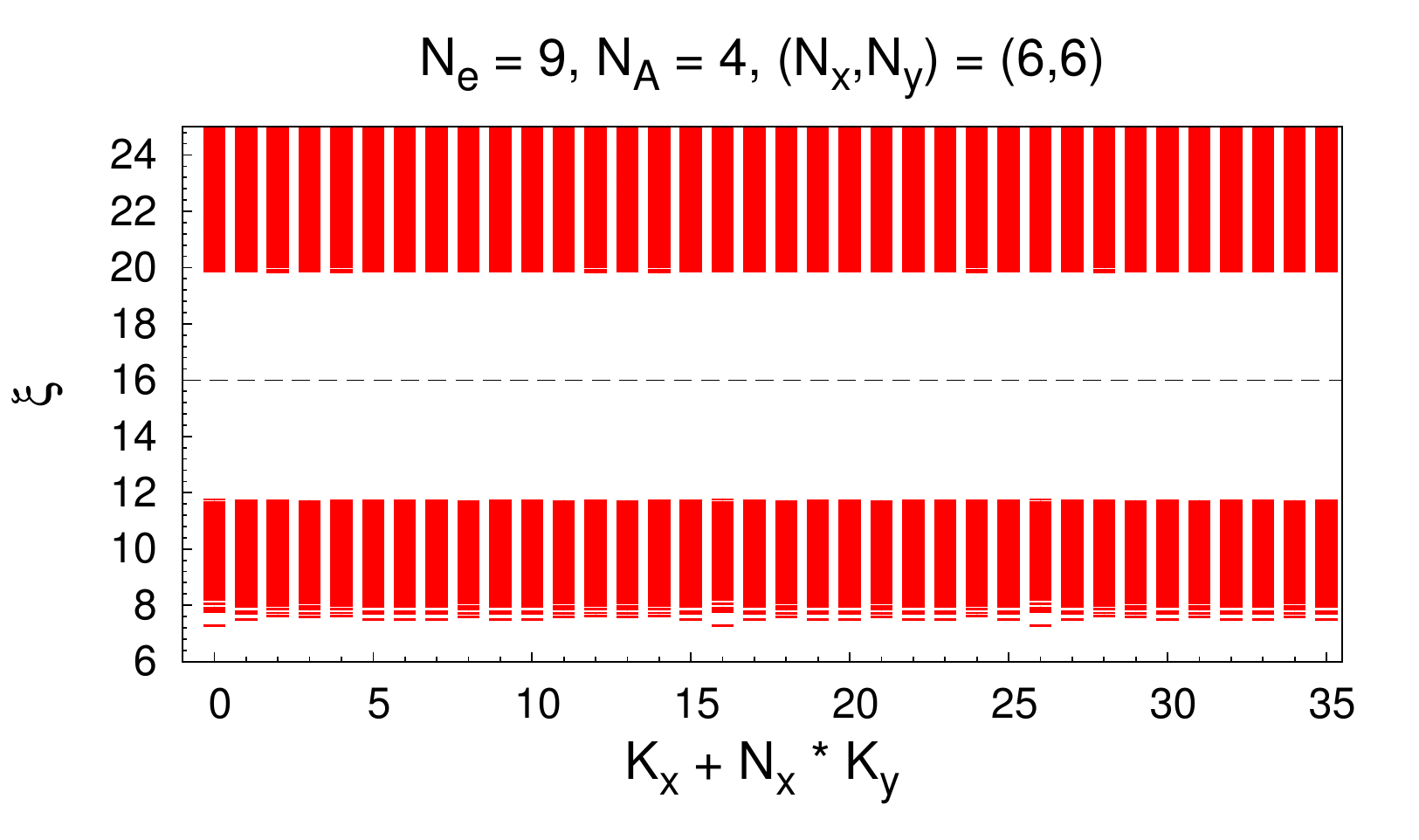}
\caption{Upper panel: PES for low energy groundstate manifold on the $2$-orbitals on triangular lattice with $C=3$ and two-body interaction for $N_e = 7$ bosons and $N_A=3$ on a $(N_x , N_y ) = (7,4)$ lattice. The number of states below the dotted line is $1428$ which is equal to the $(1,4)_1$ counting. Middle panel: PES for low energy groundstate manifold on the $2$-orbitals on triangular lattice with $C=3$ and two-body interaction for $N_e = 8$ bosons and $N_A=4$ on a $(N_x , N_y ) = (8,4)$ lattice. The number of states below the dotted line is $7112$. The $(1,4)_1$ counting gives $20$ more states per sector. Lower panel: PES for low energy groundstate manifold on the $2$-orbitals on triangular lattice with $C=3$ and two-body interaction for $N_e = 9$ bosons and $N_A=4$ on a $(N_x , N_y ) = (6,6)$ lattice. The number of states below the dotted line is $14364$.}
\label{pes_ground_state_chern3_c_3_k_1}
\end{figure}

We present the PES in Fig.~\ref{pes_ground_state_chern3_c_3_k_1} for the groundstate manifold with $N_e=7$ (upper panel), $N_e=8$ (middle panel) and $N_e=9$ (lower panel). For $N_e=7$, the PES total counting is given by the $(1,4)_1$ principle. The counting is identical to the pyrochlore case for $N_e=9$. For $N_e=8$ on the pyrochlore model, no clear entanglement gap was observed for $N_A=4$. Here the clear entanglement gap allows to deduce the counting. This latest is lower than the one of the generalized Pauli exclusion principle $(1,4)_1$, clearly showing that this phase cannot be associated to a Laughlin-like phase.

Note that for $N_e=9$, the ratio gap over spread is equal to $45370$ whereas for the pyrochlore model this ratio is equal to $58$. This makes the $C = 3$ triangular lattice the best model we have studied. As observed in Fig.~\ref{pes_ground_state_chern3_c_3_k_1}, the entanglement gaps are also bigger for this model than for the pyrochlore model.

\section{$4$-orbitals model with $C=4$}

\begin{center}
\begin{figure}[htb]
\includegraphics[width= 0.80\linewidth]{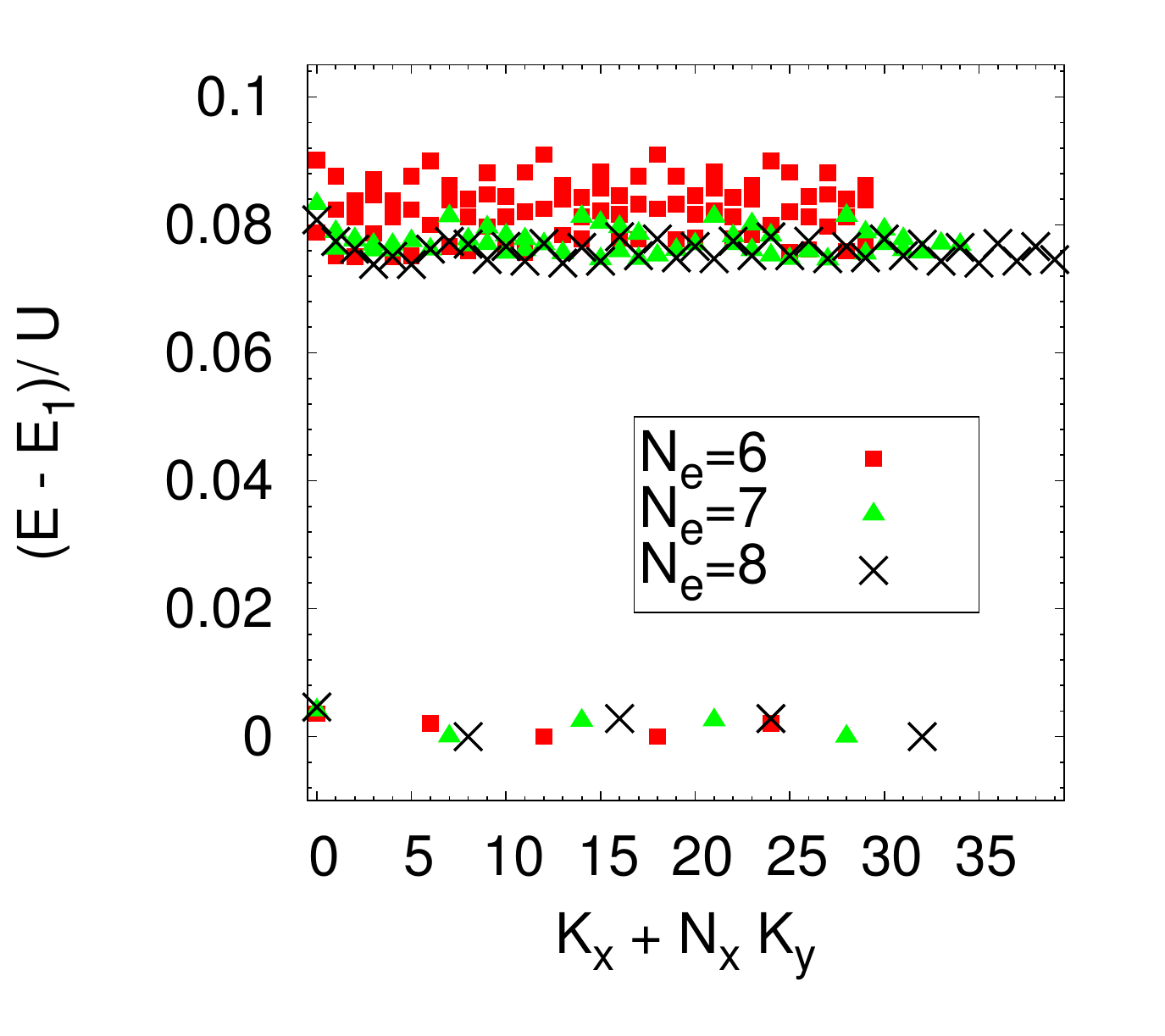}
\caption{Low energy spectra on the C-orbitals with $C=4$ and two-body interaction for $N_e = 6, 7, 8$ bosons at $\nu = \frac{1}{5}$ on a $(N_x , N_y ) = (N_e,5)$ lattice. The energies are shifted by $E_1$, the lowest energy for each system size. We observe an almost fivefold degenerate groundstate.}
\label{ground_state_norbitals}
\end{figure}
\end{center}

On the $C$-orbitals model on a square lattice~\cite{Yang-PhysRevB.86.241112}, we find convincing evidence for $C=4$ at $\nu=\frac{1}{5}$ with two-body interaction. We have used the parameters that optimize the band flatness as given in Ref.~\cite{Yang-PhysRevB.86.241112}. For this model, we have used an isotropic interaction $H_{{\rm int}, 1}=U\sum_{i} :(\sum_j \rho_{i,j})^{2}:$ where the first sum runs over all the sites whereas the second sum runs over the different orbitals on the same site. On Fig.~\ref{ground_state_norbitals}, we show the energy spectrum for the groundstate for $N_e=6,7$ and $8$ bosons. For $N_e=8$, the ratio gap over spread is equal to $15$ whereas for the pyrochlore model this ratio is equal to $6$. 

\begin{figure}[htb]
\includegraphics[width=0.98\linewidth]{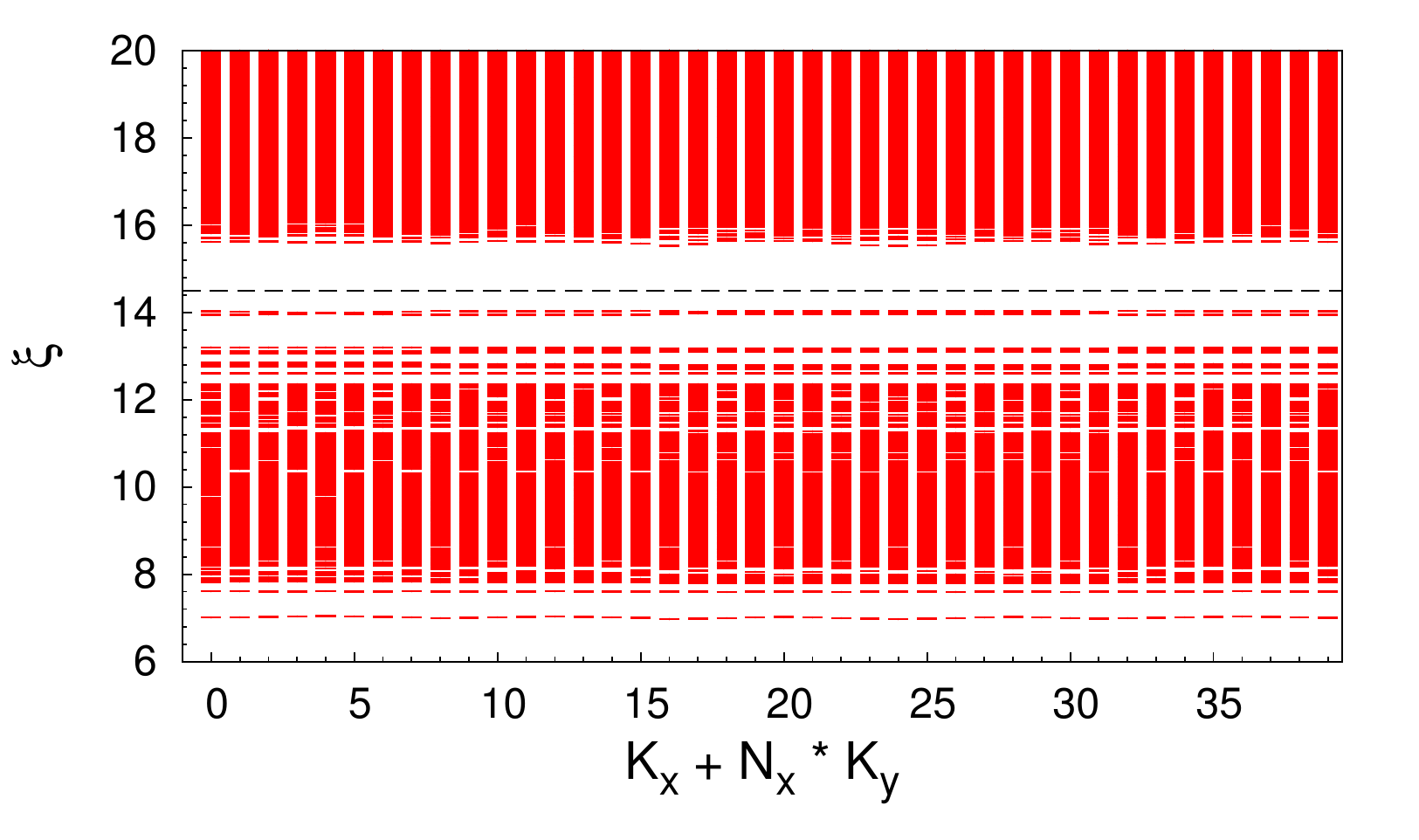}
\caption{PES for low energy groundstate manifold on the $C$-orbitals lattice with $C=4$ and two-body interaction for $N_e = 8$ bosons and $N_A=4$ on a $(N_x , N_y ) = (8,5)$ lattice. The number of states below the dotted line is $11410$. The total $(1,5)_1$ counting gives $17710$ states. The counting of Halperin state PES is $8960$ states. The number of configuration after suppressing the root configurations that involve more than $2$ particles of the same color is $15210$.}
\label{pes_ground_state_norbitals_c_2_k_2}
\end{figure}

The PES of the groundstate manifold for $N_e=8$ and $N_A=4$ is shown on Fig.~\ref{pes_ground_state_norbitals_c_2_k_2}. Interestingly, this counting does not match any of the known counting: in one hand, it is lower than the $(1,2)_3$, either complete or reduced by the configurations than involves more than three particles with the same color.  On the other hand, the counting is higher than the PES counting of the corresponding $SU(4)$ Halperin state. This example, that one should be in principle related directly to the $SU(4)$ Halperin state but does not fall into any simple explanation, might be crucial to test any prediction on the effect of dislocation on the state counting.

\end{document}